\documentclass[%
 pre,
 twocolumn,
 superscriptaddress,
 amsmath,amssymb,
 aps,
 floatfix
]{revtex4-2}

\usepackage{graphicx}
\usepackage{dcolumn}
\usepackage{bm}
\usepackage{hyperref}
\usepackage{comment}
\usepackage{upgreek}
\usepackage{notes2bib}
\usepackage{xcolor}
\usepackage{siunitx}
\usepackage{xr-hyper}

\begin{document}
\preprint{APS/123-QED}

\title{Quasi-two-dimensional bacterial swimming around pillars:\\ enhanced trapping efficiency and curvature dependence}

\author{Yuki Takaha}
 \affiliation{Department of Physics, The University of Tokyo, 7-3-1 Hongo, Tokyo 113-0033, Japan}
 \affiliation{Department of Basic Science, The University of Tokyo, 3-8-1 Komaba, Tokyo 153-8902, Japan}

\author{Daiki Nishiguchi}
 \email{nishiguchi@noneq.phys.s.u-tokyo.ac.jp}
 \affiliation{Department of Physics, The University of Tokyo, 7-3-1 Hongo, Tokyo 113-0033, Japan}
 \affiliation{PRESTO, Japan Science and Technology Agency, 4-1-8 Honcho, Saitama 332-0012, Japan}
\date{\today}

\begin{abstract}
Microswimmers exhibit more diverse behavior in quasi-two dimensions than in three dimensions.
Such behavior remains elusive due to the analytical difficulty of dealing with two parallel solid boundaries.
The existence of additional obstacles in quasi-two dimensional systems further complicates the analysis.
Combining experiments and hydrodynamic simulations, we investigate how the spatial dimension affects the interactions between microswimmers and obstacles. We fabricated microscopic pillars in quasi-two dimensions by etching glass coverslips and observed bacterial swimming among the pillars.
Bacteria got trapped around the circular pillars and the trapping efficiency increased as the quasi-two-dimensionality was increased or as the curvature of the pillars was decreased. 
Numerical simulations of the simplest situation of a confined squirmer showed anomalous increase of hydrodynamic attractions, establishing that the enhanced interaction is a universal property of quasi-two-dimensional microhydrodynamics.
We also demonstrated that the local curvature of the obstacle controls the trapping efficiency by experiments with elliptic pillars.
\end{abstract}
\maketitle

\section{Introduction}
Microswimmers such as bacteria, algae, and spermatozoa live surrounded by boundaries in natural, clinical, and experimental environments.
Bacteria often colonize epithelial surfaces of host organisms, and spermatozoa find their way to an egg inside narrow channels \cite{ishimoto2015fluid}.
The presence of such surfaces gives rise to fascinating behavior of microswimmers both at the single cell level and at the collective level. Swimming bacteria are attracted to a flat wall, which has been discussed based on the hydrodynamic interaction between the bacteria and the wall by considering mirror images of singularities of the Stokes flow \cite{berke2008hydrodynamic, blake1974fundamental} and/or on the interplay between self-propulsion and steric interactions \cite{elgeti2009selfpropelled, bianchi2017holographic}. Such attractions are of crucial importance as an initial process of biofilm formations. When a dense suspension of bacteria is confined in circular geometries \cite{wioland2013confinement, beppu2017geometry} or flown into periodic pillar arrays \cite{nishiguchi2018engineering,reinken2020organizing}, these boundaries work to rectify turbulent bacterial collective motion into ordered vortices. Several attempts have been made recently to understand the mechanism of such emergent collective behavior in bacterial turbulence from the microscopic hydrodynamic point of view not only in the bulk \cite{reinken2018derivation} but also in the presence of boundaries \cite{reinken2020organizing}.

When microswimmers are more strongly confined in a quasi-two-dimensional (quasi-2D) fluid layer between two no-slip plane walls, hydrodynamic flows around microswimmers behave differently from the bulk three-dimensional (3D) flow and thus peculiar swimming behaviors are observed. Bacteria confined in quasi-2D exhibit collective motion with long-range nematic order \cite{nishiguchi2017longrange}, which is explained by suppressed hydrodynamic instability and/or stabilizing nonequilibrium force arising from geometrical confinement \cite{maitra2018nonequilibrium}. 
Flow fields created by eukaryotic swimmers such as puller-type swimming microalgae and pusher-type dinoflagellates were also experimentally measured recently \cite{jeanneret2019confinement}. It was turned out that they have differences not only in the signs but also in the symmetries of their flows, which suggested that there remains diverse phenomena unexplored in quasi-2D active hydrodynamics. However, rigorous theoretical analysis of quasi-2D hydrodynamics between two no-slip walls has been precluded due to the necessity of considering infinite series of mirror images of higher-order singularities such as force multipoles and source multipoles \cite{liron1976stokes}, which have led to approximated or empirical calculations \cite{brotto2013hydrodynamics,cui2004anomalous,diamant2009hydrodynamic,ryan2020role, jin2021collective, houtsang2014flagella} and numerical approaches \cite{delfau2016collective}.
Therefore, unlike the mirror image technique for the 3D Stokes equations or conformal mapping for 2D ideal fluid, no easy-to-use tools for analytically calculating the interactions among multiple objects such as microswimmers and obstacles in quasi-2D have been established so far.
As these kinds of situations are frequent both in microfluidic devices and in nature, more experimental inputs and explorations are required for better understanding of the diverse life of microswimmers in quasi-2D.

In this Letter, we address the question of how pusher-type bacteria in quasi-2D interact with another boundary perpendicular to the two confining no-slip walls.
We realized this situation by fabricating microscopic pillars between two no-slip planes by a combination of photolithography and glass etching technique. 
To explore how the dimensionality affects the interactions between bacteria and no-slip pillar walls, we varied the gap width $H$ from 11.1 \si{\micro m} to as small as 1.9 \si{\micro m}, which is about twice the diameter of bacterial bodies $\sim$1 \si{\micro m}. 
We found that bacteria get more strongly trapped close to the pillar surfaces as the system became highly quasi-2D.
This was numerically supported by the increase of hydrodynamic attractive force when the gap widths were decreased.
In addition, by varying the diameter of the microscopic pillars, we found that the local curvature of the surface controls the trapping efficiency, which was directly demonstrated by an experiment with elliptic pillars.
Our findings highlight the peculiar behavior and possible control of quasi-2D ways of microswimmers' life. 

\section{Experiments}
\subsection{Method}
\begin{figure*}[tb]
\centering
\includegraphics[width=\textwidth]{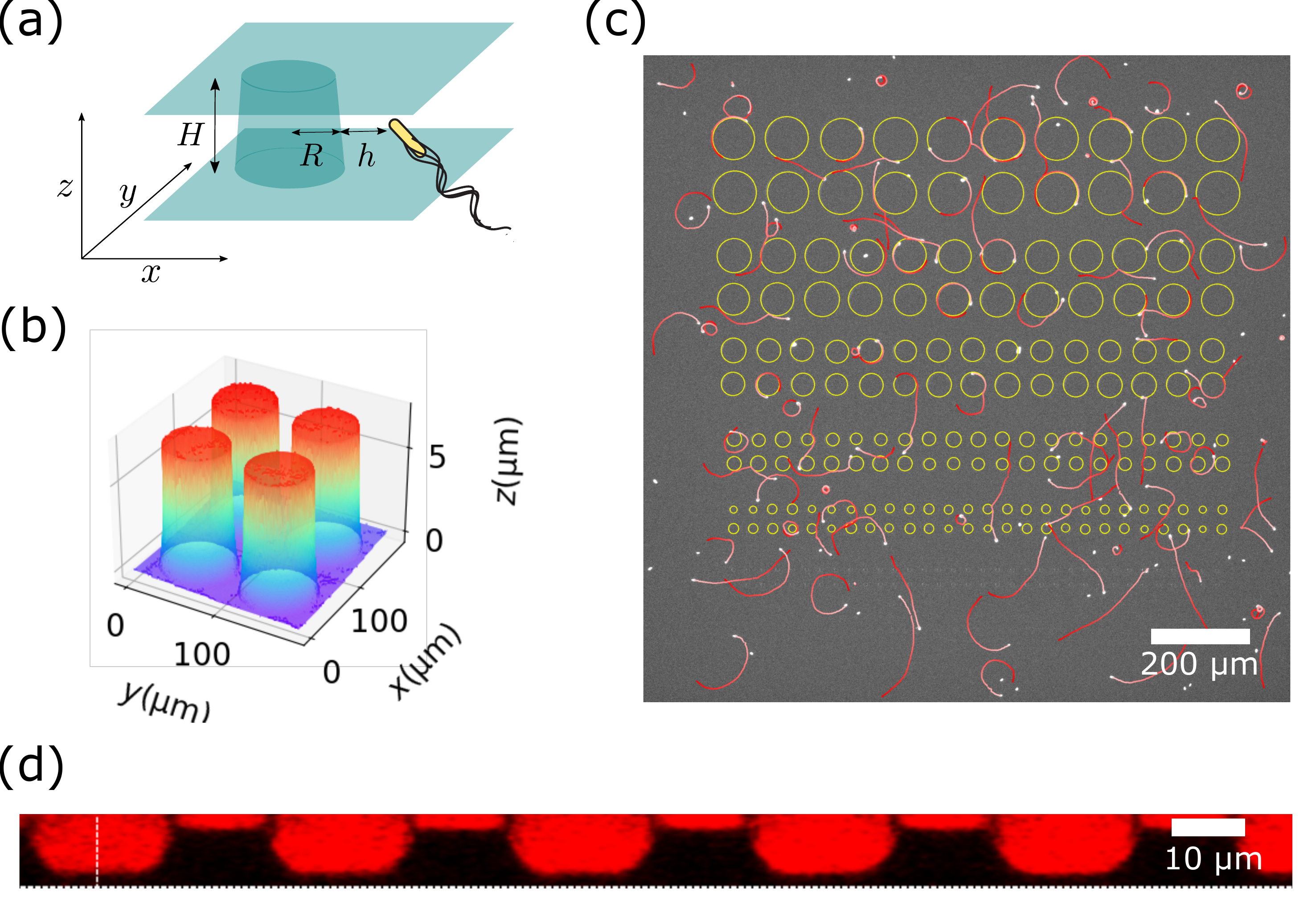}
\caption{(a) Schematic figure of our experiment. The figure is not to scale. (b) 3D surface plot of a part of the glass substrate reconstructed from confocal images. The height of the pillars is $H=6.9$ \si{\micro m}. The radii of the pillars are 38 \si{\micro m} at the bottom and 33 \si{\micro m} at the top. (c) An experimental snapshot for $H=6.9$ \si{\micro m}. Bacteria were detected as bright spots. Yellow circles are detected pillars. Red trajectories are some 10-second trajectories of the bacteria. The color changes red to white with time.(d) $xz$-slice reconstructed from the confocal images of the pillars with $R=14.3$~\si{\micro m} and $H=6.9$~\si{\micro m}. The black regions correspond to the pillars and some shadows above the pillars due to optical deflection.}
\label{fig:pillar}
\end{figure*}

In order to confine bacteria in a region as small as their body length $\sim 2$~\si{\micro m}, microscopic pillars were fabricated on a glass substrate by combining photolithography and glass etching technique (see Appendix~\ref{sec:photo}).
We arranged pillar arrays with different diameters on the same substrate (Fig.~\ref{fig:pillar}(c)), which enabled us to simultaneously obtain curvature dependence of the trapping efficiency as well as large statistics for each diameter.
In addition, constructing multiple pillars with the same diameter in a single field of view and averaging over these pillars enabled us to obtain the density distributions from dilute bacterial populations. The use of dilute bacterial suspensions also reduced unwanted collisions between bacteria which may result in forced escapes from the trapped states or sudden turns in trajectories, and thus we could focus on the single bacterium behavior.
Since the substrates were made by etching the glass coverslips in order from the surface, the pillars had larger diameters at the base than at the top (Figs.~\ref{fig:pillar}(b)(d)).
The glass substrate was washed by sonication in 20w\% aqueous soulution of alkaline detergent (Contaminon, FUJIFILM Wako Pure Chemical Corporation) and ethanol, and then rinsed with MilliQ pure water before our experiment.

We used a non-tumbling chemotactic mutant strain of {\it Escherichia coli} (RP4979) with a plasmid pZA3R-EYFP expressing yellow fluorescent protein.
{\it E. coli} were grown to the mid-exponential phase in Tryptone Broth (TB, 1 wt\% tryptone and 0.5 wt\% NaCl) with a selective antibiotic (chloramphenicol 33~\si{\micro g/mL}) at 30\si{\celsius} and then this culture was diluted 100-fold in 10~mL of fresh TB with the antibiotic at 30\si{\celsius}. 
A droplet of the bacterial suspension with L-serine was confined between the glass substrate with pillars and a glass cover slip which had been immersed in BSA solution for at least 15 minutes to prevent adhesion of bacteria to the glasses beforehand. The addition of L-serine allowed bacteria to swim actively for several minutes even in a confined environment with little oxygen\cite{lopez2015turning,douarche2009coli}.

We captured 3-minute movies at 10 frames per second at 22 \si{\degreeCelsius} by an inverted widefield fluorescence microscope (Leica, DMI6000B) equipped with an objective lens (Leica, PL APO 10x, NA 0.40 and NA 0.45 Ph for the cylindrical pillars and for the elliptic pillars respectively), a mercury lamp, and a CMOS camera (Hamamatsu, ORCA-Flash4.0 V3, 16bit, $2048\times 2048$~pixels $\simeq 1.33\times 1.33$~\si{mm^2} with the 10x objectives). 
Since this observation is at a lower magnification than, for example, those made in previous studies\cite{bianchi2017holographic,sipos2015hydrodynamic}, it is difficult to detect the exact orientation of the bacteria. Instead, the low-magnification observation allows us more reliable statistical analysis
, since the motion of many bacteria can be observed simultaneously.
We detected the trajectories of the bacteria by using an ImageJ plugin \texttt{TrackMate} \cite{tinevez2017trackmate}. The position and radius $R$ of each circular pillar were detected from the captured bright field images by using Hough transform. The obtained radius $R$ was almost the same as the radius $R_\mathrm{half}$ of the cross-section at the mid-height of the pillar that was evaluated by 3D imaging with confocal microscopy (see Appendix~\ref{sec:confocal}). The detected pillars are illustrated in Fig.~\ref{fig:pillar}(c) as circles with the radii $R$. We captured movies of experiments on three substrates: each substrate has pillars with the height of $H=1.9$~\si{\micro m}, $6.9$~\si{\micro m} or $11.1$~\si{\micro m}.
The heights of the pillars $H$ were carefully evaluated from confocal observations (see Appendix~\ref{sec:confocal} for more details of the definitions and evaluations of the geometries of pillars).

\subsection{Density distributions}
\begin{figure*}[tb]
\centering
\includegraphics[width=0.8\textwidth]{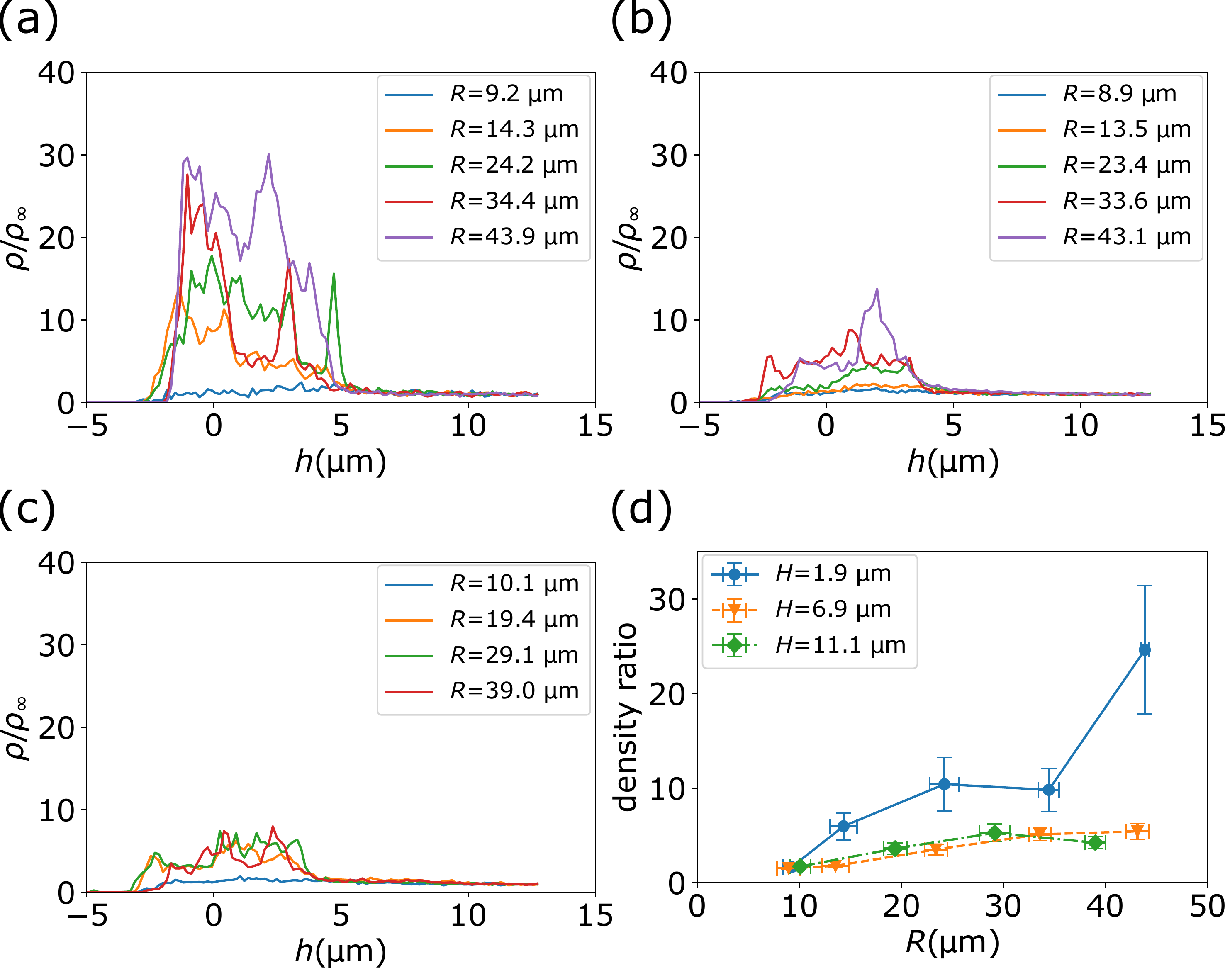}
\caption{Densities of cells around pillars with (a) $H=1.9$ \si{\micro m}, (b) $H=6.9$ \si{\micro m}, and (c) $H=11.1$ \si{\micro m}.
Bacteria were detected even at $h<0$ since the actual cross-sectional radius of the pillars varies depending on the $z$-position and can become smaller than $R$ estimated from 2D observations (see Fig.~\ref{fig:logtime}(e) and Appendix~\ref{sec:confocal}).
(d) Ratio of the densities in the nearby region to that in the far region. Error bars of the density ratio: standard errors. Error bars of the radii $R$: standard deviations.}
\label{fig:density}
\end{figure*}

In order to evaluate trapping efficiencies of the pillars, we first measured the density distributions $\rho$ as a function of the distance $h$ from the surface of each pillar (Fig.~\ref{fig:pillar}(a) and Figs.~\ref{fig:density}(a)(b)(c)). 
As the average density inevitably differs from one experiment to another due to the confinement processes and the differences in the pillar heights $H$, the density $\rho$ was normalized by the average density $\rho_\infty$ for each experiment. The average density $\rho_\infty$ was estimated from the region $10.0$ \si{\micro m}$<h<12.8$ \si{\micro m} where the density $\rho$ was well converged but still the distance $h$ was smaller than the minimal distance $d_0$ between the surfaces of adjacent pillars. 
The obtained normalized density distributions signify that the increase of the density depend on the radius of the pillars and that the trapping effect only persists up to around $h=6$~\si{\micro m} with all $H$.
We note that, although neighboring pillars are reported to influence the trapping of bacteria in the case of very small inter-pillar distance \cite{chopra2022geometric}, such influence is negligible in our case. In a previous experiment \cite{chopra2022geometric}, the pillar height $H=30$~\si{\micro m} was much larger than the inter-pillar distance $d_0=10$~\si{\micro m}, allowing the flow created by a bacterium on a pillar to reach another pillar. On the other hand, in our experiments the pillar height $H$ is always smaller than the inter-pillar distance $d_0=15$~\si{\micro m} in all the setups, which makes the flow decays fast enough to neglect the adjacent pillars in terms of hydrodynamic interactions.

Further quantification of the trapping efficiency was performed by looking at the density ratios of the nearby region $h\le6.4$~\si{\micro m} and the far region $6.4$~\si{\micro m}$<h<10.0$~\si{\micro m}.
This choice was made because the typical length of the trap was about $h=6$ \si{\micro m} regardless of the heights of the pillars (see Appendix~\ref{sec:near}).
As shown in Fig.~\ref{fig:density}(d), the pillars with larger $R$ have higher trapping efficiency than those with smaller $R$.
In previous works, such a tendency for curvature of obstacles was theoretically predicted for the trapping around a three-dimensional sphere by an analytical calculation based on the far-field approximation and mirror image technique \cite{spagnolie2015geometric} and was experimentally observed for the trapping of autocatalytic rods around solid spheres \cite{takagi2014hydrodynamic} and for bacterial trapping around pillars that were 7--40 times higher than those in our experiment \cite{sipos2015hydrodynamic}.
Importantly, in our experiments, while no difference in the trapping efficiency was observed for $H=6.9$~\si{\micro m} and $11.1$~\si{\micro m}, about up to five times more bacteria were trapped in the highly quasi-2D case of $H=1.9$~\si{\micro m}. 
In other words, the trapping efficiencies of the pillars were significantly enhanced in the highly quasi-2D setup than in the relatively 3D setups, which signifies the peculiarity of quasi-2D systems.

\subsection{Persistent length}
To evaluate the trapping efficiency also at the single cell level, we calculated persistent time and persistent length.
For simplicity, we calculated the persistent time $\tau_{R,H}$ by taking the average of the trap times at each $R$ and $H$, but it should be noted that the distribution of the trap times is skewed in the long-time direction (Figs.~\ref{fig:logtime}(a)(b)(c)).
Since the bacterial velocities depend on the heights $H$ of the confinement and can also differ from one preparation of bacterial culture to another, comparing the values of $\tau_{R,H}$ estimated from independent experiments requires careful treatment.
Therefore, in order to cancel out the dependence on the velocity, we multiplied the persistent time $\tau_{R,H}$ by the average velocity $\langle v \rangle_H$ of the bacteria in the field of view. We used this persistent length $l_{R,H}=\tau_{R,H}\langle v \rangle_H$ as a measure of trapping efficiency. 
Larger radii $R$ of pillars resulted in larger persistent lengths $l_{R,H}$ and also the pillars with $H=1.9$~\si{\micro m} had a larger persistent length than those with $H=6.9$~\si{\micro m} and $11.1$~\si{\micro m} (Fig.~\ref{fig:logtime}(d)). 
Figure~\ref{fig:logtime}(d) shows $l_{R,H}$ from two independent experiments with the substrate of $H=1.9$~\si{\micro m} but with different bacterial densities.
Both bacterial cultures were sufficiently dilute and therefore there were no discernible differences between the two experiments that may arise from the effects of bacterial collisions.
Since the actual gap widths between the top coverslip and the bottom of the substrate can vary depending on how much the cover glass was pressed down while confining the bacteria, the slight difference in the gap widths around $1.9$~\si{\micro m} significantly affects the trapping efficiency even when the same substrate was used. Such a sensitive response can also be seen in our numerical simulations that will be discussed later (Fig.~\ref{fig:squirmer}).
Nevertheless, both experiments with the substrate of $H=1.9$~\si{\micro m} consistently showed larger $l_{R,H}$ than those of larger $H$.
This again signifies the enhanced trapping efficiencies for the larger pillars and the pillars in highly quasi-2D spaces, which is consistent with the results obtained from the static information of the density distributions discussed in Fig.~\ref{fig:density}.
We note that, for small $R$, the persistent length $l_{R,H}$ does not strongly depend on $H$.
The $H$-dependence of trap efficiency appeared in situations where bacteria were no longer scattered and were trapped by the pillars.

\subsection{Chirality of trapping}
\begin{figure*}[tb]
\centering
\includegraphics[width=\textwidth]{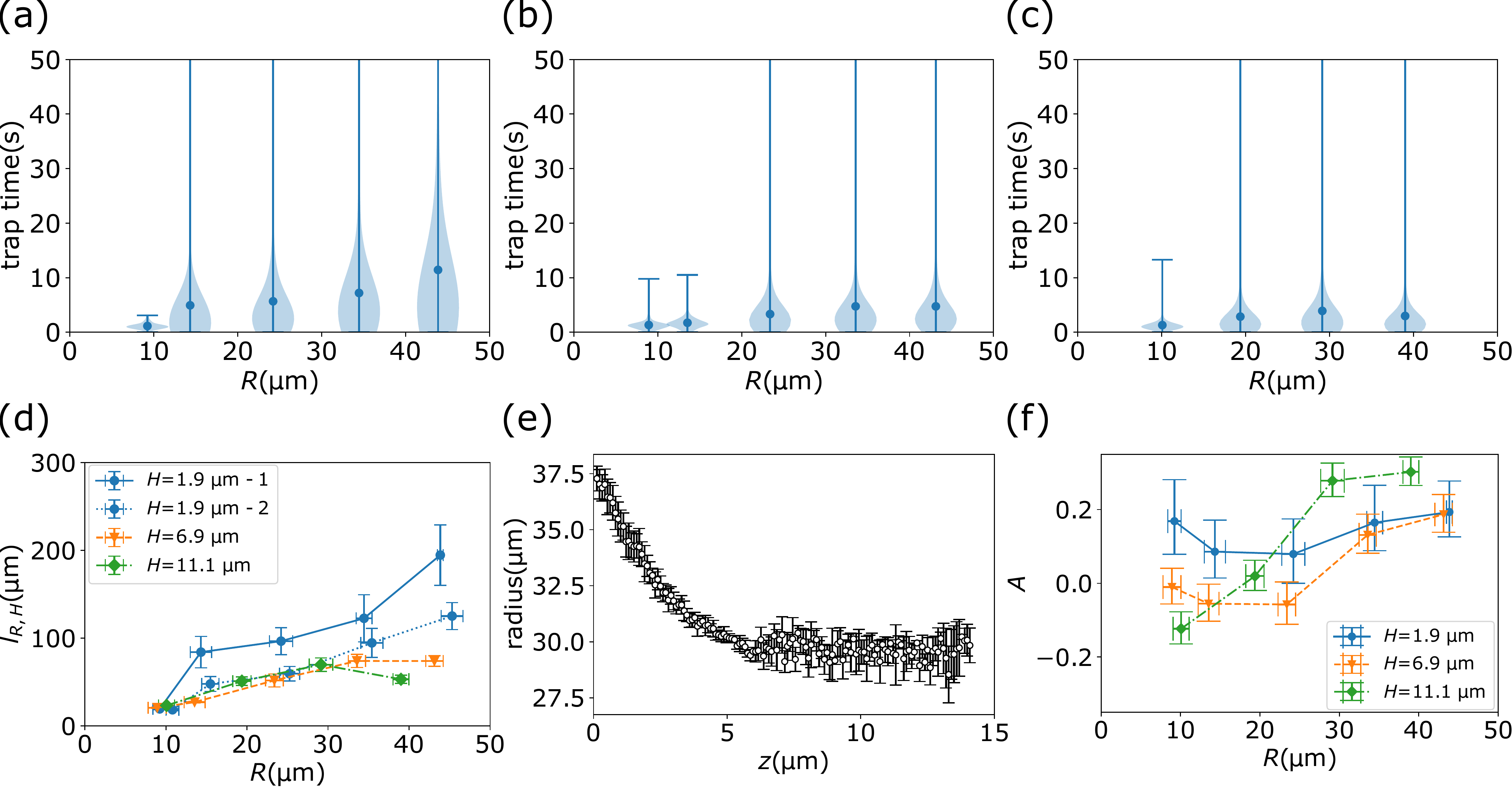}
\caption{The violin plots of the trap time with (a)$H=1.9$~\si{\micro m}, (b)$H=6.9$~\si{\micro m}, and (c)$H=11.1$~\si{\micro m}. The circular markers represent the average trap time with each $R$. (d) Persistent length $l_{R,H}$ of bacteria trapped around the pillars. The two blue plots represent the results of two independent experiments with different bacterial culture performed in different experimental areas on the same substrate with $H=1.9$~\si{\micro m}.
The plots with $H=1.9$~\si{\micro m} in Fig.~\ref{fig:density}(d)and Figs.~\ref{fig:logtime}(a)(f) correspond to the experiment labeled with $H=1.9$~\si{\micro m}-1.
Error bars: standard errors. (e) The cross-sectional radius at each $z$ position of the pillars with $H=11.1~\si{\micro m}$. Error bars: standard deviations (see Appendix~\ref{sec:confocal} for details). (f) The asymmetry parameter $A$ in the trap direction.
Error bars of $A$: 64\% confidence intervals (see Appendix~\ref{sec:asymmetry} for details).
Error bars of $R$ in (d) and (f): standard deviations.
}
\label{fig:logtime}
\end{figure*}

As swimming bacteria near a single flat wall exhibit circular trajectories due to their intrinsic chirality \cite{lauga2009}, it is natural to question whether the directions of rotations of trapped bacteria are biased or not in the quasi-2D geometry with two parallel flat walls.
To this end, we defined an asymmetry parameter $A$ in the trap direction as $A=(N_\mathrm{CW}-N_\mathrm{CCW})/({N_\mathrm{CW}+N_\mathrm{CCW})}$, where $N_\mathrm{CW}$ and $N_\mathrm{CCW}$ are the numbers of trapped trajectories in the clockwise and counterclockwise directions respectively.
The asymmetry $A$ for each pillar height and radius shown in Fig.~\ref{fig:logtime}(f) signifies that in the case of $H = 11.1$ \si{\micro m} the trapped trajectories became more biased to the clockwise direction when the radius $R$ was increased.
Since the cross-sectional radii of each pillar are larger near the bottom than that near the top (see Fig.~\ref{fig:logtime}(e)), the trapping strength should be larger near the bottom and thus the observed bacteria are more likely to be trapped near the bottom.
In addition, when viewed from above, {\it E. coli} tend to exhibit clockwise and counterclockwise trajectories near the bottom and the top respectively \cite{lauga2009}. Therefore, the observed asymmetry can be attributed to the bacteria swimming near the bottom surface.
On the other hand, almost vanishing asymmetry was observed for small radii $R$ with $H = 11.1$ \si{\micro m}.
This is because the trajectories of bacteria were more like scatterings than trapping by the small trapping efficiency.
In contrast, in the more quasi-2D cases $H = 1.9$ \si{\micro m} and $6.8$~\si{\micro m},
the asymmetry was not so strong as in the case of $H = 11.1$ \si{\micro m}.
In addition to the fact that the presence of the two walls in the quasi-2D setup compensates bacterial circular swimming and makes them swim straight \cite{swiecicki2013twodimension}, smaller differences in the cross-sectional radii of the pillars near the top and near the bottom due to the smaller etching depths $H$ diminish the differences in the trapping efficiencies, resulting in weaker asymmetry.
Thus, the asymmetry of the traps can be varied by tuning both the pillar height $H$ and radius $R$, which may be used for controlling microbial systems.

\section{Numerical experiments with a spatially fixed squirmer}
\subsection{Setup}

\begin{figure*}[tb]
\centering
\includegraphics[width=0.9\textwidth]{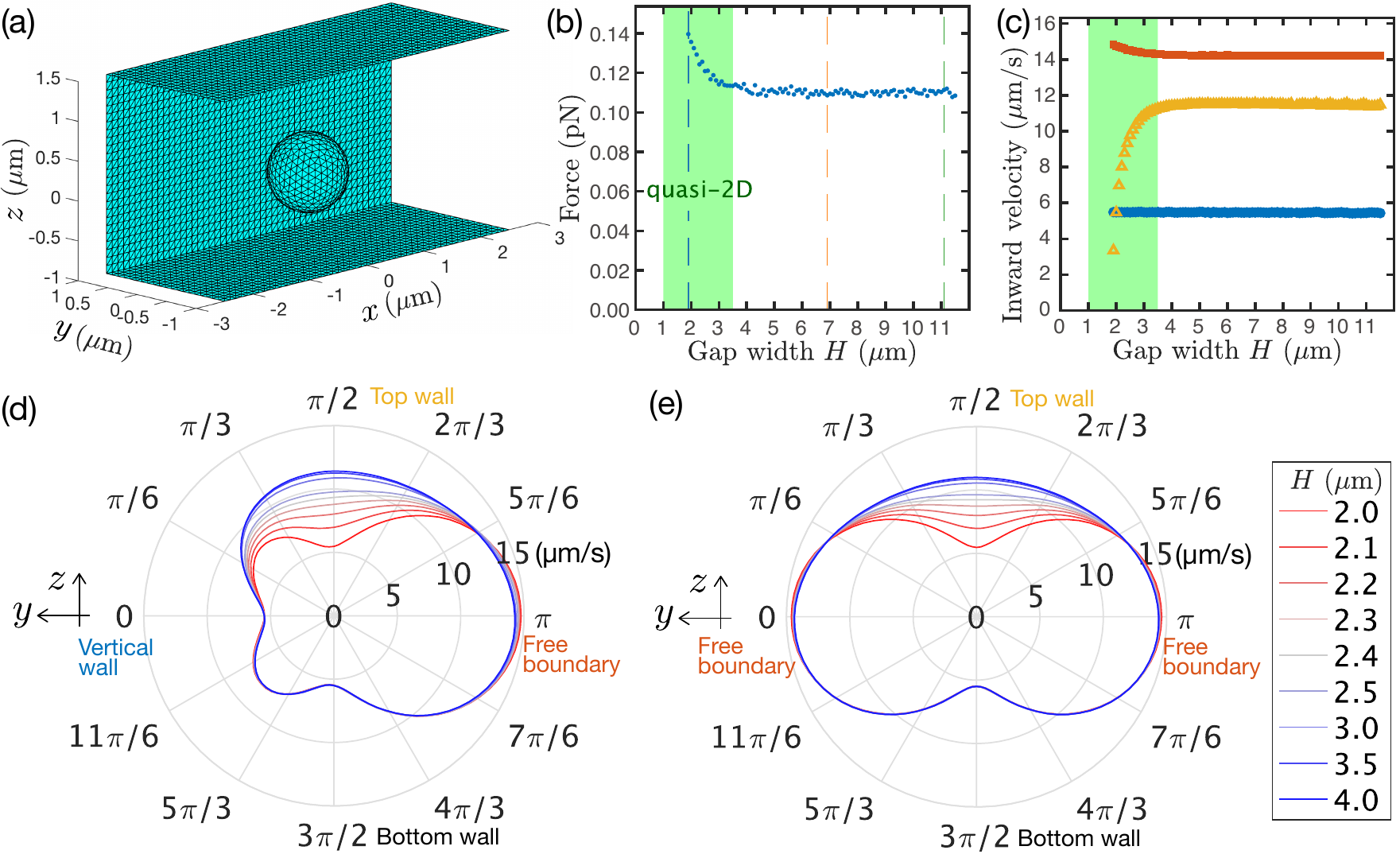}
\caption{(a) Numerical setup of the system in the case of the gap width $H=2.5$~\si{\micro m}. A spherical squirmer is placed at the origin. The three no-slip walls are shown. The depicted mesh is used for the finite element calculations (b) Force exerted on the squirmer toward the vertical wall ($+y$-direction) as a function of the gap width $H$. The attractive force is enhanced when the system becomes quasi-2D ($H\lesssim 3$~\si{\micro m}). The dashed lines correspond to experimental parameters (blue: $H=1.9$~\si{\micro m}, orange: $H=6.9$~\si{\micro m}, and green: $H=11.1$~\si{\micro m}).
(c)(d) The inward fluid velocity profile in the squirmer's equatorial plane (the $yz$-slice at $x=0$~\si{\micro m}) 0.75~\si{\micro m} (1.5 radius) away from the center of the squirmer. The velocity profile is presented in a polar plot in (d) with the origin and the direction of the angular axis flipped to be consistent with (a). The gap widths $H$ are decreased from blue lines to red lines with the same color code as in (e). In (c), the strengths of the inward velocity from the vertical wall (from +$y$-direction, blue circles), from the free boundary (from -$y$-direction, red circles), and from the top wall (from +$z$-direction, yellow triangles) are shown.
(e) Inward fluid velocity profile in the case of no vertical wall. Other setups are the same as in (d). 
}
\label{fig:squirmer}
\end{figure*}

The enhanced trapping efficiency in quasi-2D may partly be accounted for by increased hydrodynamic attraction.
To gain a universal understanding on the interactions between a microswimmer and obstacles in quasi-2D setups, we considered the simplest situation in which a model microswimmer under quasi-2D confinement was placed near a vertical flat wall.
We conducted a numerical experiment of measuring the hydrodynamic forces exerted on a squirmer spatially fixed by applying external forces with, e.g., optical tweezers in between two no-slip horizontal walls in the presence of a vertical wall representing a pillar (Fig.~\ref{fig:squirmer}(a)).

The 3D Stokes equations for this confined squirmer were solved by using a finite element method software FreeFEM++ \cite{FreeFEM}.
To focus only on the effect of dimensionality, we analyzed the interaction between the vertical flat wall at $y=+1$ \si{\micro m} and the $1$-\si{\micro m}-diameter spherical squirmer at the origin by varying the gap widths $H$ of the two horizontal planes. The bottom wall was fixed at $z=-1$ \si{\micro m} while the position of the top wall was varied from $z=10.5$~\si{\micro m} ($H=11.5$~\si{\micro m}, 3D) to $z=0.9$~\si{\micro m} ($H=1.9$~\si{\micro m}, highly quasi-2D). The parameters of the squirmer were estimated from our experimental measurements and a previous flow field measurement \cite{drescher2011fluid}: the diameter 1 \si{\micro m}, the swimming speed 20 \si{\micro m/s}, and the pusher-type swimming strength $\alpha=-3$. More detailed descriptions on the numerical methods, setups and results are given in Appendix~\ref{Appendix:NumericalExperiment}. 

\subsection{Hydrodynamic attraction}
\label{sec:HydridynamicAttraction}
Although the attractive force toward the vertical wall was observed as expected from the pusher-type flow field with inward fluid flow from all the directions at the equatorial cross section (the $yz$-plane at $x=0$~\si{\micro m}) \cite{lauga2009}, its strength was significantly enhanced when the system became highly quasi-2D with $H\lesssim 3$~\si{\micro m}.
Compared with the more 3D cases with $H\gtrsim 4$ \si{\micro m} where the attractive force stays almost constant, we observe more than $25\%$ increase of the attractive interaction. This signifies that, for this situation of the confined 1-\si{\micro m}-diameter squirmer, the effect of the upper wall sets in only for $H\lesssim 3$~\si{\micro m} where the system can regarded as truly quasi-2D. This is consistent with our experiments in which the trapping efficiencies were almost the same for $H=6.9$~\si{\micro m} and $H=11.1~\si{\micro m}$ but were significantly enhanced for $H=1.9$~\si{\micro m}.
This sharp increase of the attractive interaction around $H=2$ \si{\micro m} can be responsible for the difference in the two data sets of $H=1.9$~\si{\micro m} in Fig.~\ref{fig:logtime}(d).
Since it is difficult to control $H$ with a precision of $0.1$~\si{\micro m} in our experimental design, there were some differences in the persistent lengths $l_{R,H}$ in the two independent experiments with $H=1.9$~\si{\micro m}.

This increase of hydrodynamic attraction may be intuitively understood from the numerically obtained inward fluid velocity profiles at the equatorial cross section as a function of the gap widths $H$ (Figs.~\ref{fig:squirmer}(c)(d)).
Although in a free space without any walls the inward velocity profile around a squirmer at the equatorial cross section is isotropic, the presence of no-slip walls makes it anisotropic.
The decrease of the gap width $H$ by bringing the top wall closer to the squirmer weakens the inward flow from the top (from the $+z$-direction) toward the squirmer. This decrease of incoming flux was then compensated partly by the enhanced inward flow from the horizontal directions (the $\pm y$-directions), which is the most clearly demonstrated in the inward velocity from the free boundary as the inward flow from the vertical wall stays almost constant due to the proximity of the vertical wall.

To have a better idea of the change of the flow profiles as a function of the gap width $H$, we calculated the flow around the squirmer in the absence of the vertical wall. The vertical no-slip wall was replaced with a free boundary and the other conditions remain the same. This excludes possible effects of the vertical wall on the inward velocities on the free boundary side.
As shown in Fig.~\ref{fig:squirmer}(e), even in the absence of the vertical wall, the inward velocities from the horizontal directions with the free boundaries (the $\pm y$-directions) increased as the gap width $H$ was decreased. These results are both qualitatively and quantitatively consistent with the case with the vertical wall shown in Fig.~\ref{fig:squirmer}(d). Therefore, the increase of the inward flow from the horizontal direction, which was quantified and verified with the simplest setup, can be concluded as a universal feature of pusher-type microswimmers confined in highly quasi-2D spaces.

Just to have an idea of the magnitudes of the typical hydrodynamic force calculated here, we estimate the translational velocity of the squirmer when the external forces are turned off and the squirmer is let to move freely.
By using Stokes' law, the translational frictional coefficient for a 0.5-\si{\micro m}-radius sphere, $\gamma_t$, can be evaluated as $\gamma_t=6\pi\mu\times(0.5\;\si{\micro m})=3\pi\times10^{-3} \;\si{pN}\cdot\si{\micro m^{-1}}\cdot \si{s}$, where we assumed the shear viscosity $\mu=1\;\si{mPa\cdot s}$, which is the value of water at 20\si{\degreeCelsius}.
Therefore, the typical values of the hydrodynamic attractive force in the $+y$-direction shown in Fig.~\ref{fig:squirmer}(b), 0.11~\si{pN}, corresponds to the translational velocity of $\approx 12$~$\si{\micro m/s}$.
Thus, the hydrodynamic forces can be strong enough to explain the phenomena observed in our experiments from the viewpoint of hydrodynamics.

\subsection{Hydrodynamic torque and steric interactions}
The $z$-component of the hydrodynamic torque is responsible for reorienting the squirmer in the $xy$ plane and thus it determines whether the squirmer tries to swim toward the vertical wall ($y=+1$~\si{\micro m}) or not. However, as the $z$-component is negative (see Fig.~\ref{fig:ForceTorque}(b) in the Appendix), the reoriented squirmer tries to swim away from the vertical wall.

Although this cannot directly explain our experimental results, the steric interactions between a rod-shaped bacterium and a pillar, which is dismissed in our numerical consideration,
have been argued to play a role in keeping bacteria on the boundary \cite{elgeti2009selfpropelled, bianchi2017holographic} or guiding sperms along curved surfaces \cite{denissenko2012human, ishimoto2016simulation}. Therefore, the pillar-swimmer steric interaction should be playing an important role in stabilizing the bacterial orientation in the vicinity of the pillar and thus keeping the bacterium trapped around the pillar as previously discussed in the case of bacterial accumulations close to a flat wall \cite{elgeti2013wall}.
However, considering that the height $H$ of the pillar in our experiment was always more than twice as large as the diameter of bacterial bodies and thus the steric interaction should hardly be affected by $H$, the increased hydrodynamic forces should be playing a more dominant role in the increased trapping efficiency.

\subsection{Interpretation of the trapping mechanism}
In conclusion, the enhanced hydrodynamic attractive force between a bacterium and a pillar in quasi-2D facilitates the approach of the bacterium toward the pillar, and after it reaches the surface of the pillar then the steric interaction reorients the bacterium to keep it swim along the pillar surface by dominating the hydrodynamic torque. The hydrodynamic attraction then makes it difficult for the bacterium to escape from the nearby region of the pillar, resulting in higher trapping efficiencies in quasi-2D.

To fully confirm this story, more faithful simulations on the swimming dynamics of microswimmers in quasi-2D by using more sophisticate methods, e.g., boundary element method (BEM) \cite{ishikawa2006hydrodynamic, zhu2013low, shum2015hydrodynamicA, shum2015hydrodynamicB} or smoothed profile method (SPM) \cite{molina2013hydrodynamic, oyama2017simulations}, may be ideal for understanding how bacteria approach the pillars and for disentangling the effects of hydrodynamic and steric interactions. However, as calculations on quasi-2D setups with many boundaries are computationally quite expensive even with BEM or SPM, these are out of the scope of our current experimental study.


\section{Local curvature dependence}
\begin{figure}[tb]
\centering
\includegraphics[width=\columnwidth]{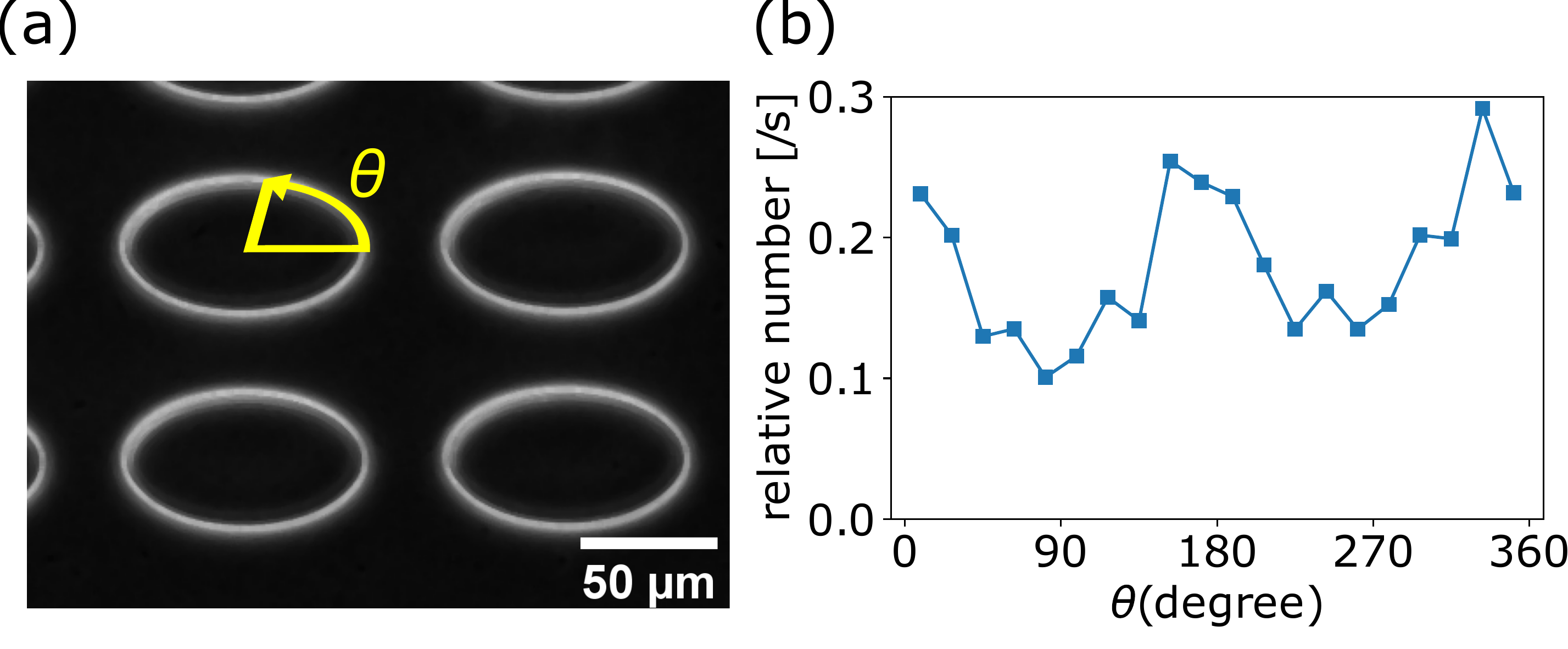}
\caption{(a) A phase contrast image of the elliptic pillars on the substrate. The lengths of semi-major axes and semi-minor axes are $52.5$ \si{\micro m} and $34.2$ \si{\micro m} respectively. (b) The number of events where the bacteria getting  out of the nearby regions devided by the number of bacteria at the angle $\theta$.
}
\label{fig:oval}
\end{figure}

Finally, to investigate whether these traps by pillars are local interactions or not, we performed additional experiments with elliptic pillars, whose nonuniform curvature prohibits the use of standard analytical techniques.
We made a glass substrate with elliptic pillars with the height $H\simeq7$~\si{\micro m} (see Fig.~\ref{fig:oval}(a)).
Since there is no function in OpenCV library of Python CV2 for directly detecting ellipses like Hough transform for detecting circles, the image analysis was performed as described in Appendix~\ref{sec:ellipse} from the captured phase contrast images.
We defined the nearby region as the internal area of a larger ellipse that was drawn by extending the semi-major axes and semi-minor axes of each elliptic pillar by $6.4$~\si{\micro m}.

As shown in Fig.~\ref{fig:oval}(b), the number  of bacteria leaving the nearby regions divided by the density was nonuniform with the peaks located at the long-axis directions.
Thus, the local trapping efficiency was larger in the direction with larger local curvature.
Similar tendency was also reported in the experiments with sedimented bimetallic-rod swimmers driven by self-electrophoresis and teardrop-shaped posts in a 3D region\cite{davieswykes2017guiding}.
Because the tendency of the higher escape rate for the larger curvatures is consistent with the experiments with the circular pillars with globally constant curvatures, the obtained nonuniform distribution directly demonstrates that the local curvature by itself can control the escape rates.
This suggests that the traps by pillars depend on local curvature of pillars and these interactions are relatively short-ranged. 
Such short-range nature of the bacterium-wall interaction in quasi-2D is also exemplified in our finite element simulation, in which the attractive force as a function of the system size $L_x$ along the $x$-direction converged at $L_x\simeq2.5$~\si{\micro m} that is comparable with the bacterial body lengths. (see Fig.~\ref{fig:squirmer}(a) and Fig.~\ref{fig:ForceXscaling} in Appendix). 

\section{Conclusion}
The quantitative differences between quasi-2D and 3D interactions found in our study is expected to lead to a better understanding of both quasi-2D far-field and near-field hydrodynamic interactions.
Our findings may trigger further quantification and theoretical development of the peculiar quasi-2D hydrodynamics
that has difficulty in rigorous theoretical analysis.
It is also important to investigate in detail the steric interactions in quasi-2D regions.
However, the analysis of the steric interactions requires high magnification that has a drawback in statistics compared with our low-magnification measurements, and thus it would be a challenging task.

From the active matter viewpoint, the pillar-swimmer interactions quantified in our experiments can also be translated into the understandings of inter-particle interactions, which gives new insight on collective behavior in lower dimensions such as bacterial long-range nematic order in quasi-2D \cite{nishiguchi2017longrange}, bacterial noncontact cohesive swimming in 2D \cite{li2017noncontact}, and colloidal swimmers under confinements \cite{nishiguchi2015mesoscopic, nishiguchi2018flagellar, poncet2020pair, iwasawa2021algebraic}.
In terms of application, the relation between the trap efficiency and the structure of the space confirmed in our experiments can serve as a design principle for active microfluidic devices using microswimmers.

\section*{Acknowledgments}
We thank Takuro Shimaya for kind technical support and discussions, Matthew S. Turner for discussions, Kazumasa A. Takeuchi for frequent discussions, and Kenta Ishimoto for valuable and stimulating discussions and critical reading of our earlier manuscript. We thank Yusuke T. Maeda and Hanna Salman for sharing plasmid DNA.
D.N. was supported by JSPS KAKENHI Grant Numbers JP19K23422, JP19H05800 and JP20K14426, and JST, PRESTO Grant Number JPMJPR21O8, Japan. 

Y.T. performed experiments with D.N. and analyzed the experimental data.  D.N. conducted numerical simulations. Both Y.T. and D.N. interpreted the results and wrote the manuscript.

\appendix

\section{Experiment}
\subsection{Washing coverslips}
First, coverslips were washed by sonication in 20w\% aqueous solution of alkaline detergent (Contaminon, FUJIFILM Wako Pure Chemical Corporation) for 20 minutes. The washed coverslips were rinsed with running tap water 15 times and then with deionized water 3 times, followed by sonication in deionized water for 2 minutes twice and then in ethanol once for 15 minutes. 
After rinsing the coverslips 3 times in deionized water, they were again washed by sonication twice in deionized water for 2 minutes and then in $8\times10^{-2}$ mol/L NaOH aqueous solution for 10 minutes.
Then, the coverslips were rinsed 3 times in deionized water and sonicated twice in deionized water for 2 minutes each. Finally, the washed coverslips were dried at 140\si{\celsius} for 40 minutes in a dry heat sterilizer.

\subsection{Microfabrication by photolithography and etching}
\label{sec:photo}
Chromium was deposited on the washed coverslips by using a thermal evaporator in order to enhance the adhesion of photoresist to the coverslips. The chromium-coated coverlips were then heated at 100 \si{\celsius} for 5 minutes on a hot plate. After the coverslips were cooled to the room temperature, 500 \si{\micro L} of the photoresist AZ 1500 (Merck Performance Materials GmbH.) was spread on the coverslips by using a spin coater (500 rpm for 5 seconds and then 2000 rpm for 30 seconds). Then the photoresist-coated coverslips were heated at 100 \si{\celsius} for one minute. 
We used a maskless aligner ($\mathrm{\mu MLA}$, Heidelberg Instruments) for exposure and wrote down the desired patterns on the photoresist.
The designed radii of pillars were 15, 20, 30, 40 and 50~\si{\micro m}. After the exposure, the photoresist were developed. We heated the coverslips at 100 \si{\celsius} for 2 minutes and then dipped them in NMD-3 2.38\% (Tokyo Ohka Kogyo Co., Ltd.), deionized water, Cr-etchant (MPM-E350, DNP Fine Chemicals Co., Ltd.) and then deionized water in order.

The exposed cover slips were etched in 50\% buffered hydrofluoric acid aqueous solution (110-BHF, Morita Chemical Industries Co., Ltd.) at 23 \si{\celsius}. The height of the fabricated pillars were controlled by tuning the etching time. After the coverslips were moved into deionized water in order to stop the etching reaction, they were rinsed in acetone, Cr-etchant and deionized water in order.

\subsection{Surface treatment}
The microfabricated substrates were washed by sonication in 20w\% aqueous solution of alkaline detergent (Contaminon, FUJIFILM Wako Pure Chemical Corporation) for 20 minutes. 
The substrates were rinsed with running tap water 15 times and then with deionized water 3 times, followed by sonication in deionized water for 2 minutes twice and then in ethanol once for 15 minutes.
After rinsing the substrates 3 times in deionized water, they were again washed by sonication twice in deionized water for 2 minutes.
After blowing off the water with a blower, a drop of 1w\% BSA (bovine serum albumin, Sigma-Aldrich) aqueous solution was put on the substrate and then we placed another coverslip without any fabrications, which would be used as a lid later in the experiment, in order to ensure that the BSA solution was spread over the whole substrate. We waited for at least 15 minutes to allow the BSA molecules to adhere to the surfaces of both the substrate and the coverslip. Right before each experiment, the substrate and the coverslip were rinsed with deionized water and the remaining water was blown off with a blower. The BSA coated surfaces of the substrate and the coverslip were used for confining the bacteria in between them subsequently.

\subsection{Confinement of {\it E. coli} into microstructures}
We used a non-tumbling chemotactic mutant strain of {\it Escherichia coli} (RP4979) with a plasmid pZA3R-EYFP expressing yellow fluorescent protein.
Bacteria taken from a frozen stock were grown overnight in Tryptone broth (T broth, 1~wt\% tryptone and 0.5~wt\% NaCl) with a selective antibiotic (chloramphenicol 33~\si{\micro g/mL}) at 30\si{\celsius} at 200~rpm and then this culture was diluted 100-fold in 10~mL of fresh T broth with the antibiotic at 30\si{\celsius} at 200~rpm.
For the experiments with the pillar heights $H=6.9$~\si{\micro m} and $11.1$~\si{\micro m}, when the optical density ($\mathrm{OD_{600}}$) of the bacterial culture reached $\mathrm{OD_{600}}=0.1$~Abs, the bacterial suspension was confined in between the microfabricated substrate and the coverslip that were precoated with BSA.
In the case of $H=1.9$~\si{\micro m}, bacteria were grown until the optical density reached $\mathrm{OD_{600}}=0.3$~Abs or $0.6$~Abs (They correspond to data 1 and 2 in Fig.~\ref{fig:logtime}(d) respectively. The plots labelled with $H=1.9$~\si{\micro m} in the other figures represent the data with $\mathrm{OD_{600}}=0.3$~Abs.) in order to enable the observation of the sufficient number of bacteria. To sustain the motility of the confined bacteria, we mixed fresh T broth containing 2.5~mol/L L-serine and the bacterial suspension in a ratio of 1:9.

Several \si{\micro L} of the prepared mixture was put on an observation area of the microfabricated substrate and then the substrate was covered with the BSA-coated coverslip. 
In order to ensure the contact of the coverslip and the top of the fabricated structures so that the bacteria do not swim above the pillars, we pressed the coverslip on the substrate.

\subsection{Characterization of fabricated pillars by confocal microscopy}
\label{sec:confocal}
Due to the etching procedure, the radii of the fabricated pillars become smaller than the original designs. Therefore, we used an inverted confocal microscope (Leica SP8) to evaluate the actual heights and radii of the fabricated pillars on the etched substrates.
We observed the pillars with the original designs of the radii 15, 20, 30, 40 and 50 \si{\micro m} on the substrates with different etching times, $40,\;120$ and $200$ minutes.
We poured 10~\si{\micro{}}mol/L rhodamin solution on the substrates and captured confocal images at every 0.1~\si{\micro m} over the range covering both the bottom of the substrates and the top of the pillars. We defined the $z$-axis in the vertical direction, and the flat surfaces of the substrates and the coverslips were placed in the $xy$-plane as in Fig.~\ref{fig:pillar}(a) in the main text.

\begin{figure*}[tb]
\centering
\includegraphics[width=\textwidth]{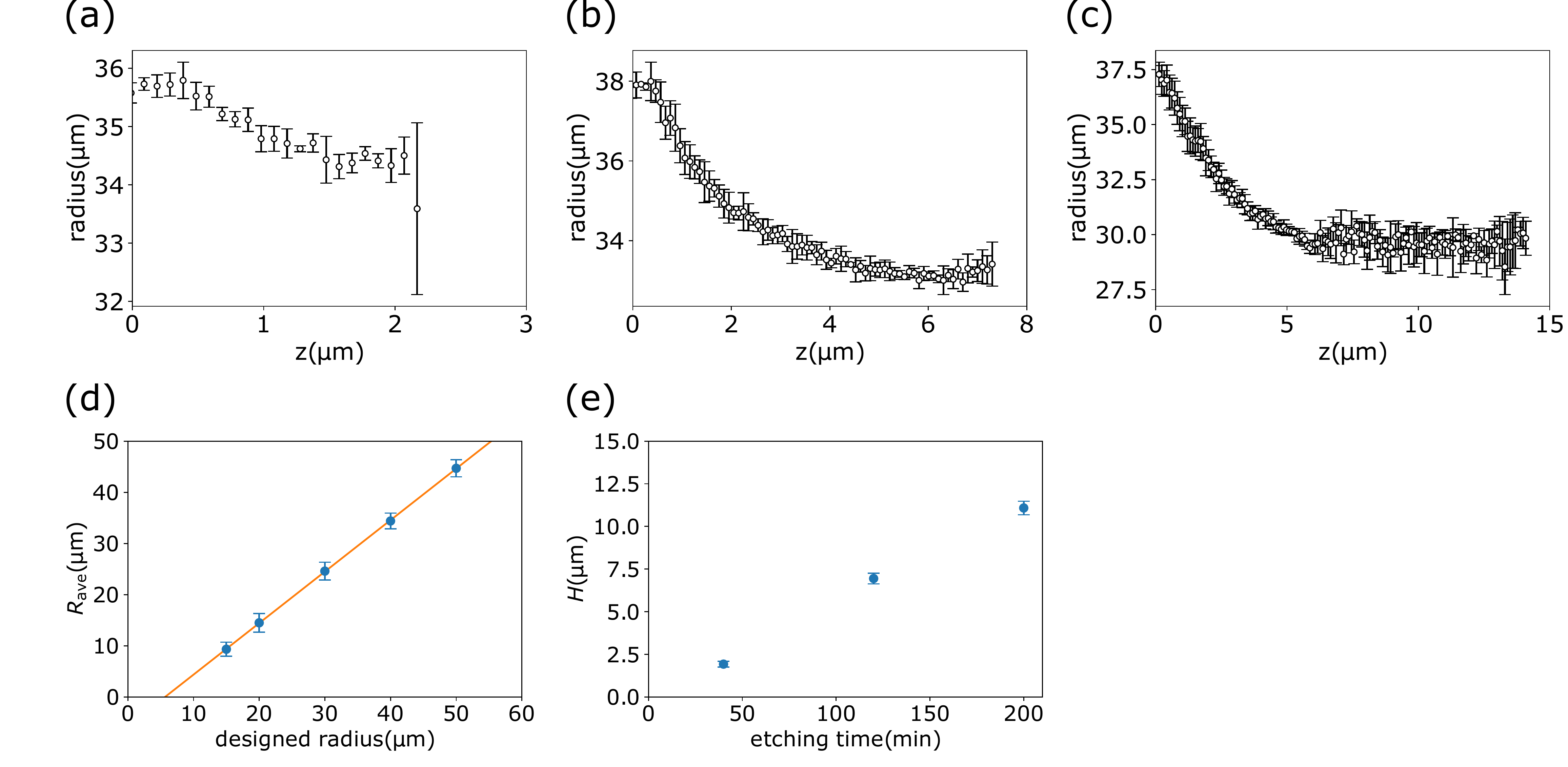}
\caption{The radius at each $z$-position of pillars with the designed radius of $40$~\si{\micro m} for (a) $H=1.9~\si{\micro m}$, (b) $H=6.9~\si{\micro m}$ and (c) $H=11.1~\si{\micro m}$. Error bars describe standard deviation at each height of 4 pillars. (d) The relation of $R_{\mathrm{ave}}$ and designed radius of pillars with $H=6.9~\si{\micro m}$. Orange line is a linear fit. Error bars: standard deviations. (e) The relation of $H$ and the etching time with $40~\si{\micro m}$ designed radius. Error bars: standard deviations.}
\label{fig:confocal}
\end{figure*}

First, we defined the centers and the radii of the pillars in the each $z$-slice of the confocal images.
Each pillar in the slice was detected as a circle by Hough transform.
Figures~\ref{fig:confocal}(a)(b)(c) show that the radii become smaller at higher $z$-positions due to the etching process.
The longer etching time resulted in the larger difference in the radii of the pillars at the top and the bottom.

The obtained positions of the centers and the radii of the pillars at each $z$-slice were then averaged to define the actual centers and radii $R_{\mathrm{ave}}$ of the pillars.
We excluded the $z$-slices where the number of the detected circles was not equal to the actual number of the pillars from the calculations of the averaged centers and the averaged radii $R_{\mathrm{ave}}$.


Next, we defined the $z$-position of the surface of the substrates.
The intensities at the largest and the smallest $z$ slices, where the whole regions are filled with the bright fluorophores and the dim glass respectively, were averaged to define the threshold intensity. This threshold was used to distinguish whether each spatial position was inside or outside of the glass substrates. The $z$-position of the surface of the substrates as functions of $x$ and $y$ were obtained by finding the $z$-positions at which the intensity exceeded the threshold value.
Since the surface thus obtained exhibited salt-and-pepper noise due to the noise in the original image, we blurred this surface by applying a median filter to obtain a smooth surface profile.

Finally, we defined the heights $H$ of pillars as the difference of the averaged $z$-positions of the bottom of the substrates and the top of the pillars. 
The $z$-positions of the bottom and the top were determined by calculating the averaged $z$-positions in the outside of circles with the radii of $\sqrt{2}R_{\mathrm{ave}}$ and the inside of circles with the radii of $R_{\mathrm{ave}}/\sqrt{2}$ whose centers coincided with those of the pillars defined above. 
The error $\sigma_{\mathrm{bottom}}$ ($\sigma_{\mathrm{top}}$) of the $z$-position of the bottom (top) was defined as the standard deviation of the $z$-positions of the pixels in the bottom (top) region defined above. 
The estimated heights $H$ of the pillars with the same designed radii of $40$ \si{\micro m} but with the different etching times were $1.9\pm0.2$ \si{\micro m}, $6.9\pm0.3$ \si{\micro m}, and $11.1\pm0.4$ \si{\micro m}, where $\pm$ means $\sqrt{\sigma_{\mathrm{top}}^2+\sigma_{\mathrm{bottom}}^2}$.
The etching rate was almost constant during $200$ minutes and was about $20$ min/\si{\micro m} (see Fig.~\ref{fig:confocal}(e)). 
Except for the case of $H=1.9$~\si{\micro m}, the height of the pillar was almost equal to the difference between the radius $R_{\mathrm{ave}}$ calculated above and the designed radius (see Fig.~\ref{fig:confocal}(d)).

Since the cross-sections of the pillars can deviate from perfect circles due to the effect of etching, we evaluated the shapes of the cross-sections at half the height of the pillars by image analysis.
We detected the edges of the cross-sections of the pillars by Canny method.
For each pillar, the distances from its center defined above to each point on the edges were averaged to define the radius of this single pillar at half height. The standard deviations $\sigma_{\mathrm{half}}$ of the distances from its center to each point on the edges were calculated for this single pillar.
Then, we calculated the ensemble average among the pillars with the same designed radii, which gave the mean radii $R_{\mathrm{half}}$. 
The standard deviations $\sigma_{\mathrm{half}}$ were smaller than $0.25$ \si{\micro m}, which is two orders of magnitude smaller than $R_{\mathrm{half}}$.
This assures that the cross-sections of pillars can be regarded as almost perfect circles.
These $\sigma_{\mathrm{half}}$ were sufficiently smaller than the differences of the radii depending on the heights.


The radii of pillars $R$ used in the main text were defined by Hough transformation of the bright field images of the experiments. 
The three definitions of the radii, $R$, $R_{\mathrm{half}}$ and $R_{\mathrm{ave}}$, gave almost the same values and consistent results.

\subsection{Definition of the density ratio}

\begin{figure*}[tb]
\centering
\includegraphics[width=\textwidth]{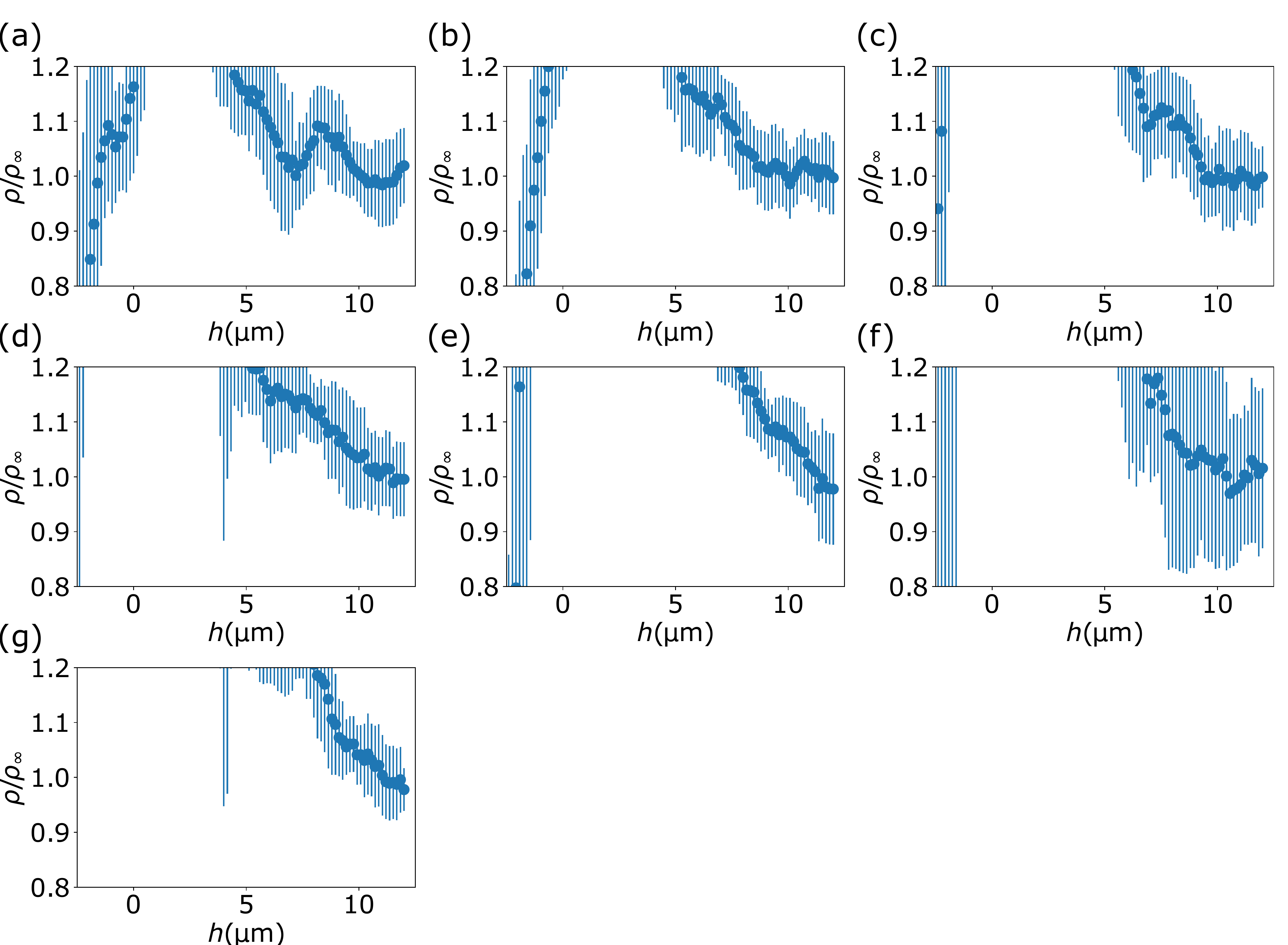}
\caption{Bacterial density and its fluctuations. (a) $H=6.9$ \si{\micro m}, $R=8.9$ \si{\micro m} (b) $H=6.9$ \si{\micro m}, $R=13.5$ \si{\micro m} (c) $H=6.9$ \si{\micro m}, $R=23.4$ \si{\micro m} (d) $H=6.9$ \si{\micro m}, $R=33.6$ \si{\micro m} (e) $H=6.9$ \si{\micro m}, $R=43.1$ \si{\micro m} (f) $H=1.9$ \si{\micro m}, $R=34.4$ \si{\micro m}, $\mathrm{OD_{600}}=0.3$~Abs, (g) $H=11.1$ \si{\micro m}, $R=29.1$ \si{\micro m}. Error bars were defined by the equation \ref{eq:sigma}.}
\label{fig:neardensity}
\end{figure*}

In defining the density in the near region, we had to pay attention to the following two points.
First, because the pillars were not perfect cylinders and also due to errors in the image analysis, bacteria could be sometimes detected in the region with $h<0$. 
Secondly, a few bacteria got sandwiched between the tops of the pillars and the cover glass used as a lid, and thus bacteria could be detected in the $h<0$ region.
Therefore, the average density of the near region was defined as the number of bacteria detected in the region $-6.4~\si{\micro m}<h<6.4~\si{\micro m}$ divided by the area of the region $0~\si{\micro m}<h<6.4~\si{\micro m}$.
The position $h=-6.4~\si{\micro m}$ (10 pixels) is a lower bound to avoid accounting for bacteria in the $h<0$ region.

\subsection{The definition of the near region}
\label{sec:near}
\begin{figure*}[tb]
\centering
\includegraphics[width=\textwidth]{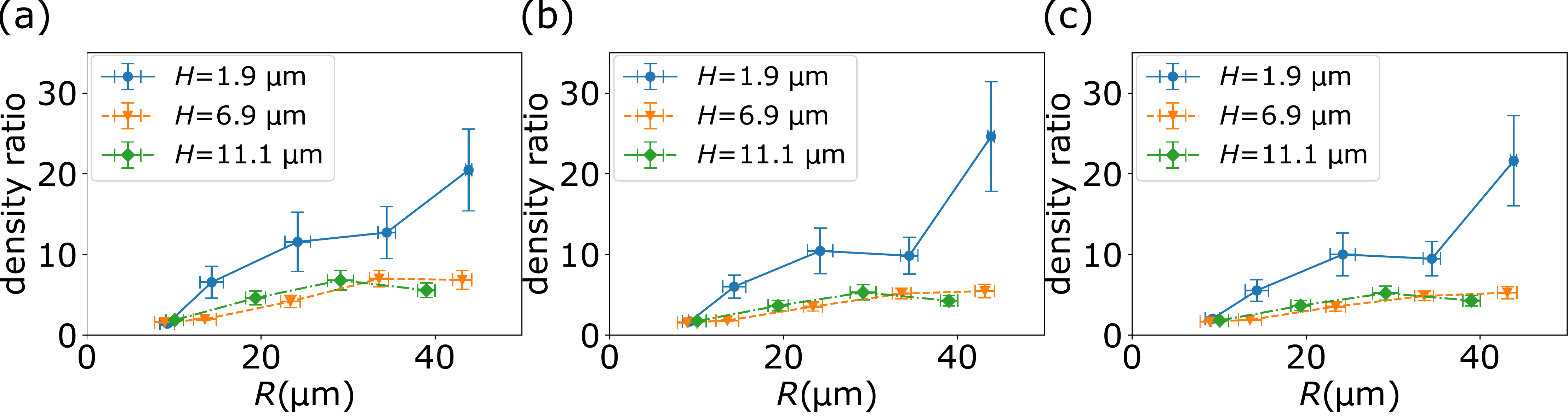}
\caption{Ratios of the density of the near region to that of the far region with different definitions. The near regions were defined as (a) $0\ \si{\micro m}<h<4.0\ \si{\micro m}$, (b) $0\ \si{\micro m}<h<6.4\ \si{\micro m}$, and (c) $0\ \si{\micro m}<h<8.0\ \si{\micro m}$. The figure (b) is used in the main text.
Error bars of density ratio: standard errors. Error bars of $R$: standard deviations.
The optical density of the bacterial culture was $\mathrm{OD_{600}}=0.3$~Abs with $H=1.9~$ \si{micro m} in these figures.}
\label{fig:nearratio}
\end{figure*}
The definition of the near region was determined by the density of bacteria and its uncertainty. First, we divided the region $-3.2\ \si{\micro m}\le h< 12.8\ \si{\micro m}$ into 100 subregions with the width of $0.16\ \si{\micro m}$ and labeled them as $I_0, I_1,\dots, I_{99}$ in the increasing order of the distance $h'_n$ from the center of the pillar to the middle point of the $n$-th subregion.
Next, we defined the average density $\rho'_n$ in the subregion $I_n$, and then we took the weighted moving average over $I_n, I_{n+1}, \dots, I_{n+10}$ to define the smoothed average density $\rho_n$ and the weighted standard deviations $\sigma_n$ at $h=0.16n-2.4\ \si{\micro m}$,
\begin{eqnarray}
 \rho_n&=&\frac{\sum_{k=n}^{n+10} (R+h'_k)\rho'_k}{\sum_{k=n}^{n+10} (R+h'_k)},\label{eq:rho}\\
 \sigma_n&=&\sqrt{\frac{\sum_{k=n}^{n+10} (R+h'_k)(\rho'_k-\rho_n)^2}{\sum_{k=n}^{n+10} (R+h'_k)}}.\label{eq:sigma}
\end{eqnarray}
Note that, farther subregions occupy larger areas than closer subregions, and the weight $R+h'_k$ is selected to properly reflect these area ratios.


Figure~\ref{fig:neardensity} shows the density$\rho$ and its uncertainty $\sigma$. In $h <5\text{--}7\ \si{\micro m}$, the density minus its uncertainty $\rho-\sigma$ is bigger than the density $\rho_{\infty}$ defined in the main text. Therefore we defined the near region as $0\ \si{\micro m}<h<6.4\ \si{\micro m}$.

We confirmed that the slight difference in the definition of the near region does not affect the behavior of our results so much (see Fig.~\ref{fig:nearratio}).



\subsection{Definition of the persistent time and length}
\begin{figure*}[tb]
\centering
\includegraphics[width=\textwidth]{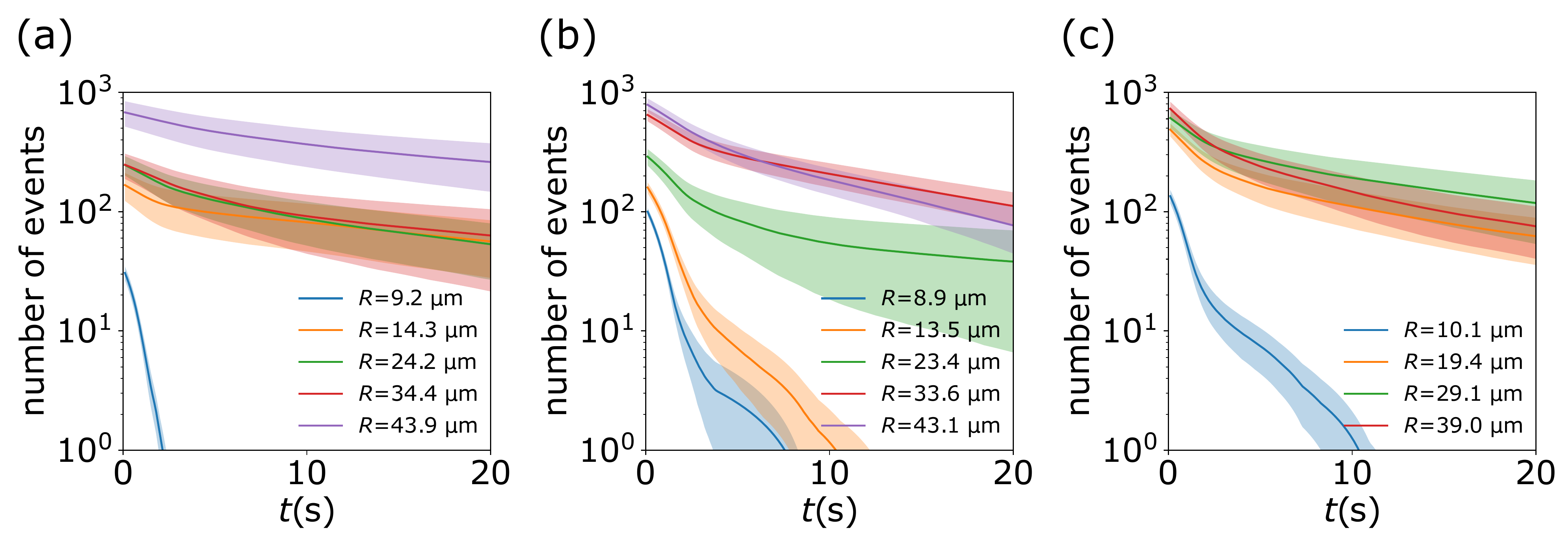}
\caption{The number of events that a bacterium keep staying in the nearby region for longer than time $t$ with (a) $H=1.9$ \si{\micro m}, $\mathrm{OD_{600}}=0.3$~Abs, (b) $H=6.9$ \si{\micro m} and (c) $H=11.1$ \si{\micro m}. Shaded error bars: standard errors.}
\label{fig:timelogs}
\end{figure*}
In the main text, the average values of the trap times are used as the definition of the persistent time.
Here we note that the persistent time cannot be well characterized by other definitions. For example, we can alternatively define the persistent time as the inverse of the decay constant if we regard the escape of a bacterium from a pillar as a Poisson process.
In reality, however, the number of events that a bacterium kept staying in the nearby region was not decaying exponentially except in the short time scale (see Fig.~\ref{fig:timelogs}), indicating that the mechanism of the escape has a memory. This memory may be associated with the orientatioal degrees of freedom of the bacteria's bodies, which play a role in trapping but were not accessible in our low-magnification observations.
Considering the complexity of the mechanism of the trapping phenomenon, the simple average of the trap times was chosen as the definition of the persistent time.

\subsection{Elliptic pillars}
\label{sec:ellipse}

Since there is no function in OpenCV library of Python CV2 for directly detecting ellipses like Hough transform for detecting circles, the image analysis was performed sequentially as follows.
The phase-contrast image was blurred and binarized and the contours in the image were detected by using the \texttt{findCounter} function in CV2.
Although the contours were detected both outside and inside the pillars , the external contours were taken as the surface of the elliptical pillars.
We determined the lengths of the major and minor axes, the positions and the orientations of the elliptic pillars by applying the \texttt{fitEllipse} function in CV2. 
Then, we defined the near region of each pillar as the internal region of an ellipse whose semi-major axis and semi-minor axis were $6.4~\si{\micro m}$ longer than those of the ellipse of the pillar.

\begin{figure*}[tb]
\centering
\includegraphics[width=\textwidth]{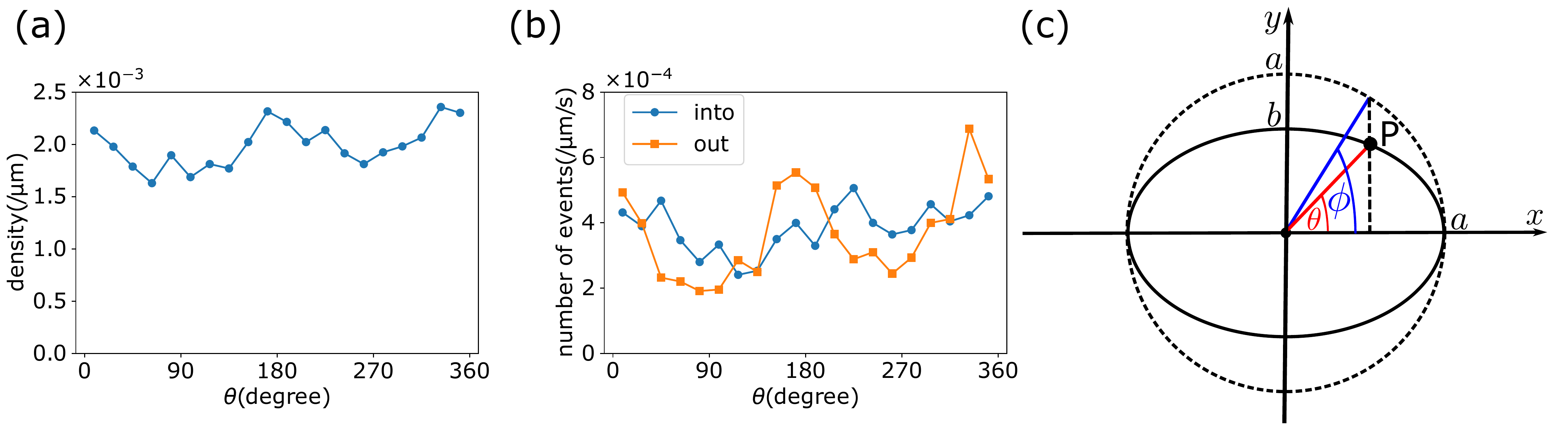}
\caption{(a) The density of the bacteria around the elliptical pillars. (b) The number of bacteria per second per arc length that got out of or into the near region. (c) The angle $\theta$ and the eccentric anomaly $\phi$ of a point $P$ on the ellipse.}
\label{fig:ellipse}
\end{figure*}

As in the case of the circular pillars, we calculated the density of bacteria around elliptical pillars and the number of events that bacteria got out of or went into their near regions, which are shown in Fig.~\ref{fig:ellipse}.
In this figure, the density in a certain direction represents the average density in the near region with an angular width of $\Delta\theta=\pi/10$ in that direction.
The average density was defined as the time-averaged number of bacteria divided by the elliptic arc length $l\Delta\theta$, which was then averaged over 115 pillars.
Here, the line element $l$ for an ellipse $x^2/a^2+y^2/b^2=1$ is given by,
\begin{eqnarray}
 l=\frac{ab(a^4\cos^2\theta+b^4\sin^2\theta)^{1/2}}{(a^2\cos^2\theta+b^2\sin^2\theta)^{3/2}}.
\end{eqnarray}
Note that the angle $\theta$ is the angle from the geometrical center of the ellipse to a point on the ellipse (see Fig.~\ref{fig:oval}(a) in the main text), not the eccentric anomaly $\phi$ which is usually used in the parametric representation of the ellipse $x=a\cos\phi,y=b\sin\phi$ (see Fig.~\ref{fig:ellipse}(c)).
The numbers of outgoing and incoming bacteria were defined as the number of events per second divided by the length of the elliptic arc length $l\Delta\theta$ and averaged over 115 pillars. 

Figure~\ref{fig:ellipse}(a) shows that the density of the bacteria showed weak, if any, nonuniformity with the peaks located at the long-axis directions. Figure~\ref{fig:ellipse}(b) shows that the number of events that the bacteria got out of the near region is anisotropic, although the number of events that the bacteria got into the near region is relatively isotropic.

\subsection{Asymmetry in the direction of rotation}
\label{sec:asymmetry}
\begin{figure*}[tb]
\centering
\includegraphics[width=0.9\textwidth]{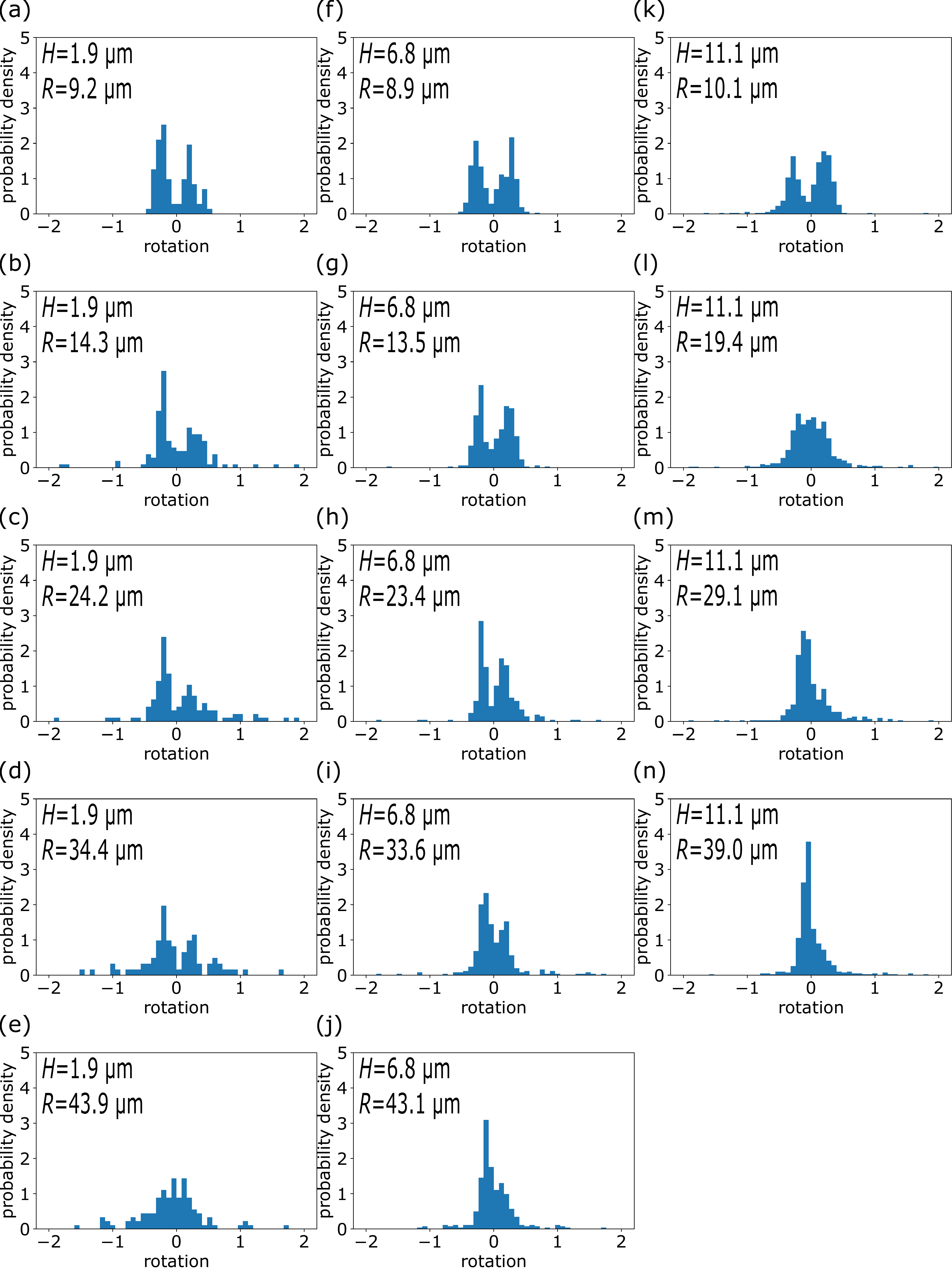}
\caption{Histograms of the rotation number for different $H$ and $R$.
The optical density of the bacterial culture was $\mathrm{OD_{600}}=0.3$~Abs with $H=1.9~$ \si{micro m}in these figures.}
\label{fig:CWCCWsupplement}
\end{figure*}

Whether the direction of bacterial rotations around each pillar is biased to a certain direction or not is a natural question from the view point of the intrinsic chirality of bacterial swimming behavior.
To this end, we obtained from the bacterial trajectories how many rotations each bacterium made around a pillar while the bacterium kept staying in the near region of the pillars.
The distribution of the angle for each pillar set with the height $H$ and the radius $R$ is shown in Fig.~\ref{fig:CWCCWsupplement}.
We defined an asymmetry parameter $A$ in the direction of rotation as in the main text, and the error of $A$ was defined by the bootstrap-like method as follows.
Suppose that the data obtained from the experiment consisted of $N$ trajectories and the number of clockwise rotations was $N_{\mathrm{CW}}$.
When a new sample consisting of $N$ trajectories is obtained by sampling again from the obtained sample with duplicates allowed, the number of clockwise rotations of the new sample follows a binomial distribution $B(N,p)$ with $p=N_{\mathrm{CW}}/N$.
Using this probability distribution, the 64\% confidence interval of $A$ is used as the errors on Fig.~\ref{fig:logtime}(f).

\section{Finite element calculation of a squirmer under confinement}
\label{Appendix:NumericalExperiment}
\subsection{Numerical setup}
To account for the increased trapping efficiency of bacteria in the quasi-two-dimensional setup in our experiment, we conducted numerical experiments to estimate the hydrodynamic forces and torques exerted on a microswimmer in the confined geometry as shown in Fig.~\ref{fig:squirmer} in the main text. Specifically, we numerically solved the 3D Stokes equations for a single pusher-type spherical squirmer whose $y$ and $z$ positions and orientation were fixed by applying external forces and torques with, e.g., optical tweezers in between two horizontal planes with no-slip boundary conditions. We placed a flat vertical wall representing a pillar in our experiment and let the squirmer to swim toward the $x$-direction. Note that we do not consider any external force in the $x$-direction so that the squirmer can only swim straight in that direction.
We estimated the strength of the hydrodynamic forces and torques as a function of the gap width $H$ between the two horizontal walls. To solve the Stokes equations, we used a finite element method software FreeFEM++ \cite{FreeFEM}.

The detailed numerical setup is as follows. A spherical squirmer with the diameter of 1 \si{\micro m} is fixed at the origin and tries to swim in the $+x$-direction by giving the surface slip velocity,
\begin{equation}
    v_s(\theta) = \frac{3}{2}U\left( \sin{\theta}+\frac{\alpha}{2}\sin{2\theta} \right),
\end{equation}
along the angular unit vector,
\begin{equation}
\bm{\mathrm{e}}_\theta =\frac{1}{\sqrt{x^2+y^2+z^2}}\left(
    \begin{array}{c}
      -\sqrt{y^2+z^2} \\
      x\cos{\arctan{\frac{z}{y}}} \\
      x\sin{\arctan{\frac{z}{y}}} 
    \end{array}
  \right),
\end{equation}
where $\theta=\arctan{\frac{\sqrt{y^2+z^2}}{x}}$ is the angle from the $x$ axis, $U$ is the swimming speed of the squirmer in a bulk fluid without any walls, and $\alpha$ represents the strength of the force dipole of the swimmer. Here, the parameters were set to be $U=20\;\si{\micro m/s}$ and $\alpha=-3$ so that they reproduce our experimental measurements of bacterial swimming speed and the previous measurements of the bacterial flow field \cite{drescher2011fluid}.
We placed the lower horizontal wall at $z=-1\;\si{\micro m}$, and we changed the height $H$ of the system by changing the position of the upper horizontal wall from $z=+0.9\;\si{\micro m}$ to $z=+10.5\;\si{\micro m}$, which corresponds to our experiments with the varying gap widths $H$ from $1.9\;\si{\micro m}$ to $11.6\;\si{\micro m}$. The vertical wall is placed at $y=+1$~\si{\micro m} with the no-slip conditions, and other boundaries at $y=-1\;\si{\micro m}$, $x=+2.5\;\si{\micro m}$, and $x=-2.5\;\si{\micro m}$ are free boundaries without any constraints on the velocity. The actual mesh on the solid boundaries used for tetrahedralization in the finite element method is shown in Fig.~\ref{fig:squirmer}(a). The system size was restricted by the requirement of the RAM $>800$GB for the larger systems even with the use of relatively large sizes of the mesh and tetrahedra, which resulted in erroneous variations seen in the data of Fig.~\ref{fig:squirmer}(b).

\begin{figure}[tb]
\includegraphics[width=\columnwidth]{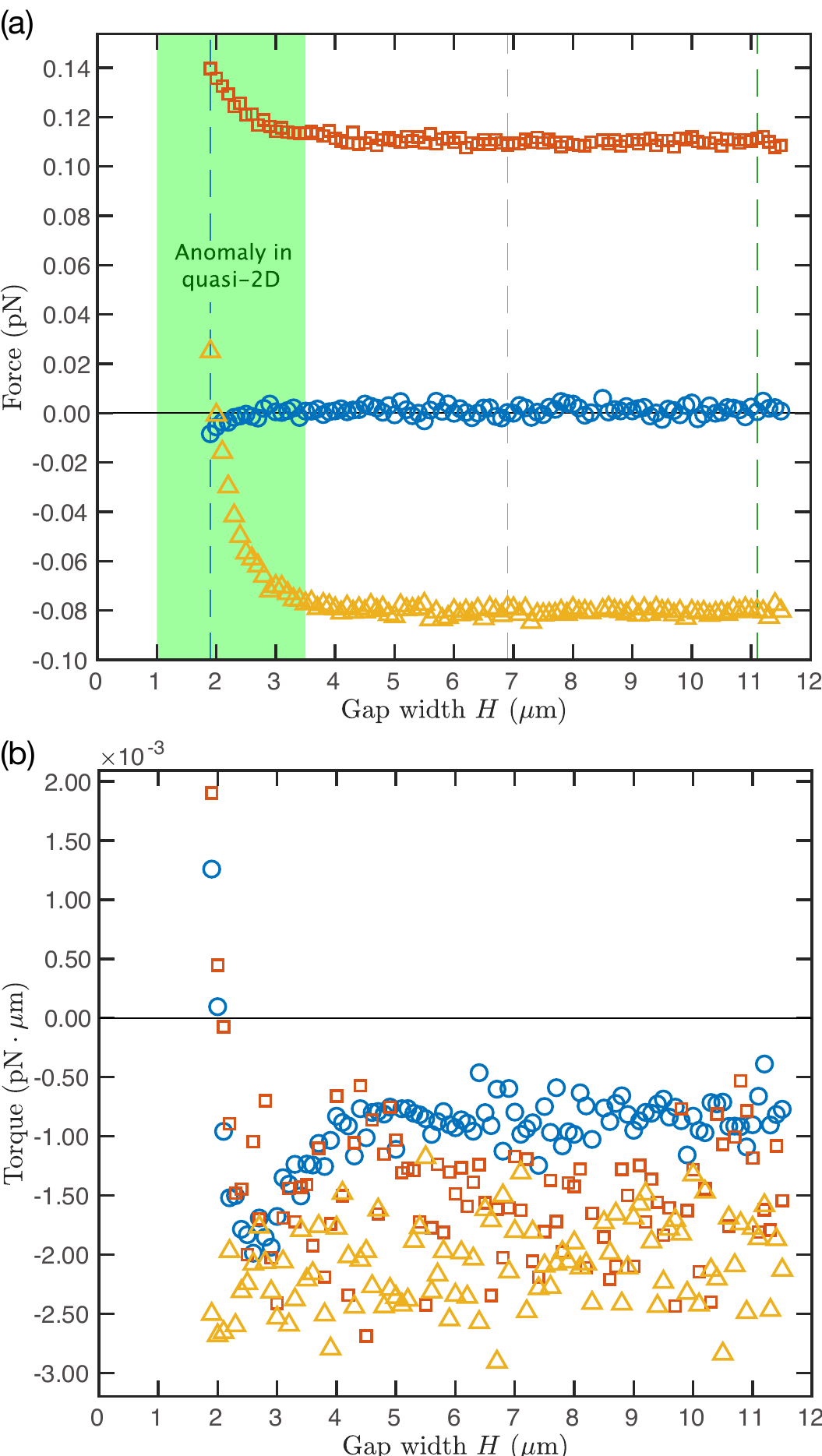}
\caption{\label{fig:ForceTorque} The numerically obtained hydrodynamic (a) force and (b) torque exerted on a squirmer as functions of the gap width $H$ of the two confining planes. Blue circles: $x$ component. Red squares: $y$ component. Yellow triangles: $z$ component. The three dahsed lines in (a) correspond to experimental parameters: blue for $H=1.9$~\si{\micro m}, orange for $H=6.9$~\si{\micro m}, and green for $H=11.1$~\si{\micro m}.
}
\end{figure}

\begin{figure}[tb]
\includegraphics[width=\columnwidth]{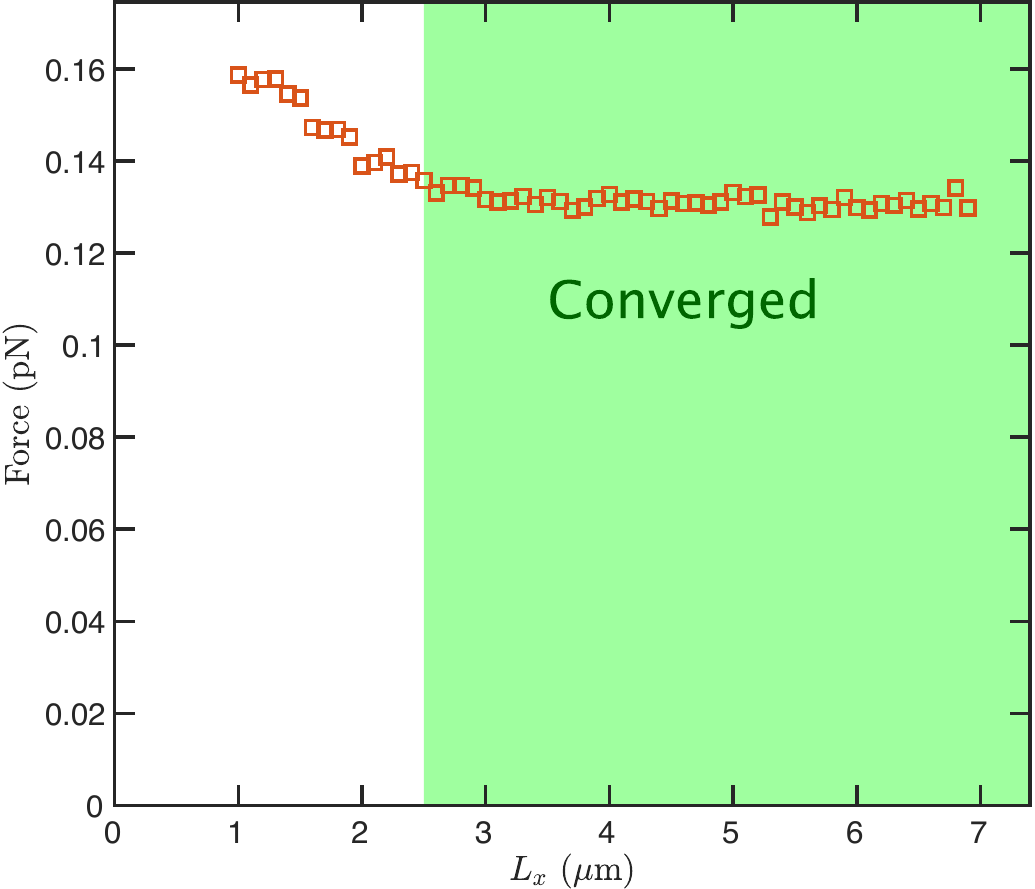}
\caption{\label{fig:ForceXscaling} 
Numerical resutls on the hydrodynamic force acting on the squirmer toward the vertical wall ($+y$-direction) as a function of the system size in the $x$-direction $L_x$. The force can be regarded converged for $L_x>2.5\;\si{\micro m}$.
}
\end{figure}

To solve the equation in the laboratory frame, the actual swimming velocity $U_0 \bm{\mathrm{e}}_x$ of the squirmer in the presence of the boundaries should be subtracted from the surface slip velocity, where ${\mathrm{e}}_x$ is the unit vector along the $x$-axis. The actual swimming velocity is usually calculated to satisfy both the force-free and torque-free conditions of a squirmer, but here, as we consider a situation in which the squirmer is both positionally and orientationally fixed except for the $x$-direction, the force free and the torque free conditions do not hold. Thus, we set $U_0 \approx U$ as a first approximation. This approximation will later be justified by the fact that the obtained hydrodynamic force in the $x$-direction is almost zero within the range of numerical errors (see Fig.~\ref{fig:ForceTorque}(a)).
Therefore, the given slip velocity at $(x,y,z)$ on the surface of the squirmer is,
\begin{equation}
 v_s(\theta)\bm{\mathrm{e}}_\theta-U\bm{\mathrm{e}}_x=\frac{ v_s(\theta)}{\sqrt{x^2+y^2+z^2}}\left(
    \begin{array}{c}
      -\sqrt{y^2+z^2} \\
      x\cos{\arctan{\frac{z}{y}}} \\
      x\sin{\arctan{\frac{z}{y}}} 
    \end{array}
  \right)
  -\left(
    \begin{array}{c}
      U \\
      0 \\
      0 
    \end{array}
  \right).
\end{equation}
After obtaining the full velocity field $\bm{u}(\bm{r})$, we calculated the hydrodynamic force $\bm{F}$ and torque $\bm{T}$ exerted on the squirmer by integrating the stress field on the surface of the squirmer as,
\begin{eqnarray}
 \bm{F}&=&\int_S \sigma\cdot \bm{n} dS,\\
 \bm{T}&=&\int_S \bm{r}\times \sigma\cdot \bm{n} dS,\\
 \sigma&=&\mu(\nabla\bm{u}+\nabla\bm{u}^T)-pI,
\end{eqnarray}
where $S$ represents the surface of the squirmer, $\sigma$ is the stress tensor, $\bm{n}$ is the unit normal vector of the squirmer surface pointing outward, $dS$ is the surface element, $\mu$ is the shear viscosity of the fluid, $p$ is the static pressure, and $I$ is the identity matrix. 
The indices $i$ and $j$ represent the spatial coordinates. We set $\mu=1\;\si{mPa\cdot s}$, which is the value of water at 20\si{\degreeCelsius}. All the $x$, $y$, and $z$ components of the force and the torque are shown in Fig.~\ref{fig:ForceTorque}.
In Figs.~\ref{fig:squirmer}(b) and \ref{fig:ForceXscaling}, only the $y$ components of $\bm{F}$ are shown for visibility.

\subsection{Strengths of hydrodynamic forces and torques}


Here we evaluate the translational and angular velocities of the squirmer when the external forces and torques are turned off and the squirmer is free to move as we did in the main text (Sec.~\ref{sec:HydridynamicAttraction}).
Stokes' law gives the translational and rotational frictional coefficient for a 0.5-\si{\micro m}-radius sphere, $\gamma_t$ and $\gamma_r$ respectively, as $\gamma_t=6\pi\mu\times(0.5\;\si{\micro m})=3\pi\times10^{-3} \;\si{pN}\cdot\si{\micro m^{-1}}\cdot \si{s}$ and $\gamma_r=8\pi\mu\times(0.5\;\si{\micro m})^3=\pi\times10^{-3}\;\si{pN}\cdot\si{\micro m}\cdot \si{s}$.
By using these formulae, the typical values of the hydrodynamic attractive force in the $+y$-direction and the $z$-component of the torque, 0.11~\si{pN} and $-2\times 10^{-3}$~$\si{pN}\cdot\si{\micro m}$, correspond to the translational velocity of $\approx 12$~$\si{\micro m/s}$ and the angular velocity of $\approx -\frac{2}{\pi}$~\si{rad/s} respectively, both of which are strong enough for hydrodynamics to play a role.

\subsection{System size dependence of the numerical results}
As the choice of the system size can affect systematically the numerical results, we have checked the system-size dependence of the hydrodynamic attractive force by performing the calculations by changing the system size in $x$-directions with the top wall fixed at $z=+1\;\si{\micro m}$. The other setups are the same. The vertical wall was placed at $y=+1\;\si{\micro m}$, and the free boundaries are at $y=-1\;\si{\micro m}$, $x=+L_x$ and $x=-L_x$. We increased $L_x$ from $1\;\si{\micro m}$ to $10\;\si{\micro m}$, which confirmed that the estimated hydrodynamic force can be assumed to be converged above $L_x=2.5\;\si{\micro m}$. This is the reason of our choice of $L_x=2.5\;\si{\micro m}$ in our simulations with variable gap widths.


\section{Movie description}
\begin{description}
\item[Supplemental Movie 1] (1\_H1.9um\_WholeView\_speed3x.avi)\\
The whole field of view of the circular pillar experiment with $H=1.9$~\si{\micro m} and $\mathrm{OD_{600}}=0.3$~Abs. The movie shows the first one minute out of the experimentally captured three-minute movie used for the analysis. The spatial resolution is scaled down to $1024\times 1024$ from the original $2048 \time 2048$ and the temporal resolution is lowered by only showing every three frames. The playback speed is three times the real speed. The contrast of the images is enhanced to make the pillars visible.

\item[Supplemental Movie 2] (2\_H6.9um\_WholeView\_speed3x.avi)\\
The whole field of view of the circular pillar experiment with $H=6.9$~\si{\micro m}. The movie shows the first one minute out of the experimentally captured three-minute movie used for the analysis. The spatial resolution is scaled down to $1024\times 1024$ from the original $2048 \time 2048$ and the temporal resolution is lowered by only showing every three frames. The playback speed is three times the real speed. The contrast of the images is enhanced to make the pillars visible.

\item[Supplemental Movie 3] (3\_H11.1um\_WholeView\_speed3x.avi)\\
The whole field of view of the circular pillar experiment with $H=11.1$~\si{\micro m}. The movie shows the first one minute out of the experimentally captured three-minute movie used for the analysis. The spatial resolution is scaled down to $1024\times 1024$ from the original $2048 \time 2048$ and the temporal resolution is lowered by only showing every three frames. The playback speed is three times the real speed. The contrast of the images is enhanced to make the pillars visible.

\item[Supplemental Movie 4] (4\_H1.9um\_Zoom\_speed3x.avi)\\
The same experiment as Supplemental Movie 1 with a cropped view of $360\times 400$ pixels view for the first 1--600 frames out of the total 1800 frames. The original spatial and temporal resolutions are kept to ensure the visibility. The playback speed is three times the real speed. The contrast of the images is enhanced to make the pillars visible.

\item[Supplemental Movie 5] (5\_H6.9um\_Zoom\_speed3x.avi)\\
The same experiment as Supplemental Movie 2 with a cropped view of $360\times 400$ pixels view for the first 1--600 frames out of the total 1800 frames. The original spatial and temporal resolutions are kept to ensure the visibility. The playback speed is three times the real speed. The contrast of the images is enhanced to make the pillars visible.

\item[Supplemental Movie 6] (6\_H11.1um\_Zoom\_speed3x.avi)\\
The same experiment as Supplemental Movie 3 with a cropped view of $360\times 400$ pixels view for the first 1--600 frames out of the total 1800 frames. The original spatial and temporal resolutions are kept to ensure the visibility. The playback speed is three times the real speed. The contrast of the images is enhanced to make the pillars visible.

\item[Supplemental Movie 7]
(7\_H6.9um\_Tracking\_speed3x.avi)\\
The same experiment as Supplemental Movie 2 with tracking performed by using TrackMate for the total 1800 frames. Detected bacteria are enclosed with magenta circles and their trajectories are represented with red lines. The spatial resolution is scaled down to $512\times 512$ from the original $2048 \time 2048$. The original temporal resolutions are kept to ensure the visibility. The playback speed is three times the real speed. The contrast of the images is enhanced to make the pillars visible.

\item[Supplemental Movie 8] (8\_Ellipse\_H7um\_WholeView\_speed3x.avi)\\
The whole field of view of the elliptice pillar experiment with $H\simeq 7$~\si{\micro m}. The movie shows the first one minute out of the experimentally captured three-minute movie used for the analysis. The spatial resolution is scaled down to $1024\times 1024$ from the original $2048 \time 2048$ and the temporal resolution is lowered by only showing every three frames. The playback speed is three times the real speed. The contrast of the images is enhanced to make the pillars visible.

\item[Supplemental Movie 9] (9\_Ellipse\_H7um\_Zoom\_speed3x.avi)\\
The same experiment as Supplemental Movie 8 with a cropped view of $380\times 300$ pixels view for the first 1--600 frames out of the total 1800 frames. The original spatial and temporal resolutions are kept to ensure the visibility. The playback speed is three times the real speed. The contrast of the images is enhanced to make the pillars visible.

\end{description}

\bibliography{ref}

\begin{thebibliography}{47}%
\makeatletter
\providecommand \@ifxundefined [1]{%
 \@ifx{#1\undefined}
}%
\providecommand \@ifnum [1]{%
 \ifnum #1\expandafter \@firstoftwo
 \else \expandafter \@secondoftwo
 \fi
}%
\providecommand \@ifx [1]{%
 \ifx #1\expandafter \@firstoftwo
 \else \expandafter \@secondoftwo
 \fi
}%
\providecommand \natexlab [1]{#1}%
\providecommand \enquote  [1]{``#1''}%
\providecommand \bibnamefont  [1]{#1}%
\providecommand \bibfnamefont [1]{#1}%
\providecommand \citenamefont [1]{#1}%
\providecommand \href@noop [0]{\@secondoftwo}%
\providecommand \href [0]{\begingroup \@sanitize@url \@href}%
\providecommand \@href[1]{\@@startlink{#1}\@@href}%
\providecommand \@@href[1]{\endgroup#1\@@endlink}%
\providecommand \@sanitize@url [0]{\catcode `\\12\catcode `\$12\catcode
  `\&12\catcode `\#12\catcode `\^12\catcode `\_12\catcode `\%12\relax}%
\providecommand \@@startlink[1]{}%
\providecommand \@@endlink[0]{}%
\providecommand \url  [0]{\begingroup\@sanitize@url \@url }%
\providecommand \@url [1]{\endgroup\@href {#1}{\urlprefix }}%
\providecommand \urlprefix  [0]{URL }%
\providecommand \Eprint [0]{\href }%
\providecommand \doibase [0]{https://doi.org/}%
\providecommand \selectlanguage [0]{\@gobble}%
\providecommand \bibinfo  [0]{\@secondoftwo}%
\providecommand \bibfield  [0]{\@secondoftwo}%
\providecommand \translation [1]{[#1]}%
\providecommand \BibitemOpen [0]{}%
\providecommand \bibitemStop [0]{}%
\providecommand \bibitemNoStop [0]{.\EOS\space}%
\providecommand \EOS [0]{\spacefactor3000\relax}%
\providecommand \BibitemShut  [1]{\csname bibitem#1\endcsname}%
\let\auto@bib@innerbib\@empty
\bibitem [{\citenamefont {Ishimoto}\ and\ \citenamefont
  {Gaffney}(2015)}]{ishimoto2015fluid}%
  \BibitemOpen
  \bibfield  {author} {\bibinfo {author} {\bibfnamefont {K.}~\bibnamefont
  {Ishimoto}}\ and\ \bibinfo {author} {\bibfnamefont {E.~A.}\ \bibnamefont
  {Gaffney}},\ }\bibfield  {title} {\bibinfo {title} {Fluid flow and sperm
  guidance: a simulation study of hydrodynamic sperm rheotaxis},\ }\href
  {https://doi.org/10.1098/rsif.2015.0172} {\bibfield  {journal} {\bibinfo
  {journal} {Journal of The Royal Society Interface}\ }\textbf {\bibinfo
  {volume} {12}},\ \bibinfo {pages} {20150172} (\bibinfo {year}
  {2015})}\BibitemShut {NoStop}%
\bibitem [{\citenamefont {Berke}\ \emph {et~al.}(2008)\citenamefont {Berke},
  \citenamefont {Turner}, \citenamefont {Berg},\ and\ \citenamefont
  {Lauga}}]{berke2008hydrodynamic}%
  \BibitemOpen
  \bibfield  {author} {\bibinfo {author} {\bibfnamefont {A.~P.}\ \bibnamefont
  {Berke}}, \bibinfo {author} {\bibfnamefont {L.}~\bibnamefont {Turner}},
  \bibinfo {author} {\bibfnamefont {H.~C.}\ \bibnamefont {Berg}},\ and\
  \bibinfo {author} {\bibfnamefont {E.}~\bibnamefont {Lauga}},\ }\bibfield
  {title} {\bibinfo {title} {Hydrodynamic attraction of swimming microorganisms
  by surfaces},\ }\href {https://doi.org/10.1103/PhysRevLett.101.038102}
  {\bibfield  {journal} {\bibinfo  {journal} {Physical Review Letters}\
  }\textbf {\bibinfo {volume} {101}},\ \bibinfo {pages} {038102} (\bibinfo
  {year} {2008})}\BibitemShut {NoStop}%
\bibitem [{\citenamefont {Blake}\ and\ \citenamefont
  {Chwang}(1974)}]{blake1974fundamental}%
  \BibitemOpen
  \bibfield  {author} {\bibinfo {author} {\bibfnamefont {J.~R.}\ \bibnamefont
  {Blake}}\ and\ \bibinfo {author} {\bibfnamefont {A.~T.}\ \bibnamefont
  {Chwang}},\ }\bibfield  {title} {\bibinfo {title} {Fundamental singularities
  of viscous flow},\ }\href {https://doi.org/10.1007/BF02353701} {\bibfield
  {journal} {\bibinfo  {journal} {Journal of Engineering Mathematics}\ }\textbf
  {\bibinfo {volume} {8}},\ \bibinfo {pages} {23} (\bibinfo {year}
  {1974})}\BibitemShut {NoStop}%
\bibitem [{\citenamefont {Elgeti}\ and\ \citenamefont
  {Gompper}(2009)}]{elgeti2009selfpropelled}%
  \BibitemOpen
  \bibfield  {author} {\bibinfo {author} {\bibfnamefont {J.}~\bibnamefont
  {Elgeti}}\ and\ \bibinfo {author} {\bibfnamefont {G.}~\bibnamefont
  {Gompper}},\ }\bibfield  {title} {\bibinfo {title} {Self-propelled rods near
  surfaces},\ }\href {https://doi.org/10.1209/0295-5075/85/38002} {\bibfield
  {journal} {\bibinfo  {journal} {EPL (Europhysics Letters)}\ }\textbf
  {\bibinfo {volume} {85}},\ \bibinfo {pages} {38002} (\bibinfo {year}
  {2009})}\BibitemShut {NoStop}%
\bibitem [{\citenamefont {Bianchi}\ \emph {et~al.}(2017)\citenamefont
  {Bianchi}, \citenamefont {Saglimbeni},\ and\ \citenamefont
  {Di~Leonardo}}]{bianchi2017holographic}%
  \BibitemOpen
  \bibfield  {author} {\bibinfo {author} {\bibfnamefont {S.}~\bibnamefont
  {Bianchi}}, \bibinfo {author} {\bibfnamefont {F.}~\bibnamefont
  {Saglimbeni}},\ and\ \bibinfo {author} {\bibfnamefont {R.}~\bibnamefont
  {Di~Leonardo}},\ }\bibfield  {title} {\bibinfo {title} {Holographic imaging
  reveals the mechanism of wall entrapment in swimming bacteria},\ }\href
  {https://doi.org/10.1103/PhysRevX.7.011010} {\bibfield  {journal} {\bibinfo
  {journal} {Physical Review X}\ }\textbf {\bibinfo {volume} {7}},\ \bibinfo
  {pages} {011010} (\bibinfo {year} {2017})}\BibitemShut {NoStop}%
\bibitem [{\citenamefont {Wioland}\ \emph {et~al.}(2013)\citenamefont
  {Wioland}, \citenamefont {Woodhouse}, \citenamefont {Dunkel}, \citenamefont
  {Kessler},\ and\ \citenamefont {Goldstein}}]{wioland2013confinement}%
  \BibitemOpen
  \bibfield  {author} {\bibinfo {author} {\bibfnamefont {H.}~\bibnamefont
  {Wioland}}, \bibinfo {author} {\bibfnamefont {F.~G.}\ \bibnamefont
  {Woodhouse}}, \bibinfo {author} {\bibfnamefont {J.}~\bibnamefont {Dunkel}},
  \bibinfo {author} {\bibfnamefont {J.~O.}\ \bibnamefont {Kessler}},\ and\
  \bibinfo {author} {\bibfnamefont {R.~E.}\ \bibnamefont {Goldstein}},\
  }\bibfield  {title} {\bibinfo {title} {Confinement stabilizes a bacterial
  suspension into a spiral vortex},\ }\href
  {https://doi.org/10.1103/PhysRevLett.110.268102} {\bibfield  {journal}
  {\bibinfo  {journal} {Physical Review Letters}\ }\textbf {\bibinfo {volume}
  {110}},\ \bibinfo {pages} {268102} (\bibinfo {year} {2013})}\BibitemShut
  {NoStop}%
\bibitem [{\citenamefont {Beppu}\ \emph {et~al.}(2017)\citenamefont {Beppu},
  \citenamefont {Izri}, \citenamefont {Gohya}, \citenamefont {Eto},
  \citenamefont {Ichikawa},\ and\ \citenamefont {Maeda}}]{beppu2017geometry}%
  \BibitemOpen
  \bibfield  {author} {\bibinfo {author} {\bibfnamefont {K.}~\bibnamefont
  {Beppu}}, \bibinfo {author} {\bibfnamefont {Z.}~\bibnamefont {Izri}},
  \bibinfo {author} {\bibfnamefont {J.}~\bibnamefont {Gohya}}, \bibinfo
  {author} {\bibfnamefont {K.}~\bibnamefont {Eto}}, \bibinfo {author}
  {\bibfnamefont {M.}~\bibnamefont {Ichikawa}},\ and\ \bibinfo {author}
  {\bibfnamefont {Y.~T.}\ \bibnamefont {Maeda}},\ }\bibfield  {title} {\bibinfo
  {title} {Geometry-driven collective ordering of bacterial vortices},\ }\href
  {https://doi.org/10.1039/C7SM00999B} {\bibfield  {journal} {\bibinfo
  {journal} {Soft Matter}\ }\textbf {\bibinfo {volume} {13}},\ \bibinfo {pages}
  {5038} (\bibinfo {year} {2017})}\BibitemShut {NoStop}%
\bibitem [{\citenamefont {Nishiguchi}\ \emph {et~al.}(2018)\citenamefont
  {Nishiguchi}, \citenamefont {Aranson}, \citenamefont {Snezhko},\ and\
  \citenamefont {Sokolov}}]{nishiguchi2018engineering}%
  \BibitemOpen
  \bibfield  {author} {\bibinfo {author} {\bibfnamefont {D.}~\bibnamefont
  {Nishiguchi}}, \bibinfo {author} {\bibfnamefont {I.~S.}\ \bibnamefont
  {Aranson}}, \bibinfo {author} {\bibfnamefont {A.}~\bibnamefont {Snezhko}},\
  and\ \bibinfo {author} {\bibfnamefont {A.}~\bibnamefont {Sokolov}},\
  }\bibfield  {title} {\bibinfo {title} {Engineering bacterial vortex lattice
  via direct laser lithography},\ }\href
  {https://doi.org/10.1038/s41467-018-06842-6} {\bibfield  {journal} {\bibinfo
  {journal} {Nature Communications}\ }\textbf {\bibinfo {volume} {9}},\
  \bibinfo {pages} {4486} (\bibinfo {year} {2018})}\BibitemShut {NoStop}%
\bibitem [{\citenamefont {Reinken}\ \emph {et~al.}(2020)\citenamefont
  {Reinken}, \citenamefont {Nishiguchi}, \citenamefont {Heidenreich},
  \citenamefont {Sokolov}, \citenamefont {B{\"a}r}, \citenamefont {Klapp},\
  and\ \citenamefont {Aranson}}]{reinken2020organizing}%
  \BibitemOpen
  \bibfield  {author} {\bibinfo {author} {\bibfnamefont {H.}~\bibnamefont
  {Reinken}}, \bibinfo {author} {\bibfnamefont {D.}~\bibnamefont {Nishiguchi}},
  \bibinfo {author} {\bibfnamefont {S.}~\bibnamefont {Heidenreich}}, \bibinfo
  {author} {\bibfnamefont {A.}~\bibnamefont {Sokolov}}, \bibinfo {author}
  {\bibfnamefont {M.}~\bibnamefont {B{\"a}r}}, \bibinfo {author} {\bibfnamefont
  {S.~H.}\ \bibnamefont {Klapp}},\ and\ \bibinfo {author} {\bibfnamefont
  {I.~S.}\ \bibnamefont {Aranson}},\ }\bibfield  {title} {\bibinfo {title}
  {Organizing bacterial vortex lattices by periodic obstacle arrays},\ }\href
  {https://doi.org/10.1038/s42005-020-0337-z} {\bibfield  {journal} {\bibinfo
  {journal} {Communications Physics}\ }\textbf {\bibinfo {volume} {3}},\
  \bibinfo {pages} {76} (\bibinfo {year} {2020})}\BibitemShut {NoStop}%
\bibitem [{\citenamefont {Reinken}\ \emph {et~al.}(2018)\citenamefont
  {Reinken}, \citenamefont {Klapp}, \citenamefont {B{\"a}r},\ and\
  \citenamefont {Heidenreich}}]{reinken2018derivation}%
  \BibitemOpen
  \bibfield  {author} {\bibinfo {author} {\bibfnamefont {H.}~\bibnamefont
  {Reinken}}, \bibinfo {author} {\bibfnamefont {S.~H.~L.}\ \bibnamefont
  {Klapp}}, \bibinfo {author} {\bibfnamefont {M.}~\bibnamefont {B{\"a}r}},\
  and\ \bibinfo {author} {\bibfnamefont {S.}~\bibnamefont {Heidenreich}},\
  }\bibfield  {title} {\bibinfo {title} {Derivation of a hydrodynamic theory
  for mesoscale dynamics in microswimmer suspensions},\ }\href
  {https://doi.org/10.1103/PhysRevE.97.022613} {\bibfield  {journal} {\bibinfo
  {journal} {Physical Review E}\ }\textbf {\bibinfo {volume} {97}},\ \bibinfo
  {pages} {022613} (\bibinfo {year} {2018})}\BibitemShut {NoStop}%
\bibitem [{\citenamefont {Nishiguchi}\ \emph {et~al.}(2017)\citenamefont
  {Nishiguchi}, \citenamefont {Nagai}, \citenamefont {Chat{\'e}},\ and\
  \citenamefont {Sano}}]{nishiguchi2017longrange}%
  \BibitemOpen
  \bibfield  {author} {\bibinfo {author} {\bibfnamefont {D.}~\bibnamefont
  {Nishiguchi}}, \bibinfo {author} {\bibfnamefont {K.~H.}\ \bibnamefont
  {Nagai}}, \bibinfo {author} {\bibfnamefont {H.}~\bibnamefont {Chat{\'e}}},\
  and\ \bibinfo {author} {\bibfnamefont {M.}~\bibnamefont {Sano}},\ }\bibfield
  {title} {\bibinfo {title} {Long-range nematic order and anomalous
  fluctuations in suspensions of swimming filamentous bacteria},\ }\href
  {https://doi.org/10.1103/PhysRevE.95.020601} {\bibfield  {journal} {\bibinfo
  {journal} {Physical Review E}\ }\textbf {\bibinfo {volume} {95}},\ \bibinfo
  {pages} {020601(R)} (\bibinfo {year} {2017})}\BibitemShut {NoStop}%
\bibitem [{\citenamefont {Maitra}\ \emph {et~al.}(2018)\citenamefont {Maitra},
  \citenamefont {Srivastava}, \citenamefont {Marchetti}, \citenamefont
  {Lintuvuori}, \citenamefont {Ramaswamy},\ and\ \citenamefont
  {Lenz}}]{maitra2018nonequilibrium}%
  \BibitemOpen
  \bibfield  {author} {\bibinfo {author} {\bibfnamefont {A.}~\bibnamefont
  {Maitra}}, \bibinfo {author} {\bibfnamefont {P.}~\bibnamefont {Srivastava}},
  \bibinfo {author} {\bibfnamefont {M.~C.}\ \bibnamefont {Marchetti}}, \bibinfo
  {author} {\bibfnamefont {J.~S.}\ \bibnamefont {Lintuvuori}}, \bibinfo
  {author} {\bibfnamefont {S.}~\bibnamefont {Ramaswamy}},\ and\ \bibinfo
  {author} {\bibfnamefont {M.}~\bibnamefont {Lenz}},\ }\bibfield  {title}
  {\bibinfo {title} {A nonequilibrium force can stabilize 2{D} active
  nematics},\ }\href {https://doi.org/10.1073/pnas.1720607115} {\bibfield
  {journal} {\bibinfo  {journal} {Proceedings of the National Academy of
  Sciences}\ }\textbf {\bibinfo {volume} {115}},\ \bibinfo {pages} {6934}
  (\bibinfo {year} {2018})}\BibitemShut {NoStop}%
\bibitem [{\citenamefont {Jeanneret}\ \emph {et~al.}(2019)\citenamefont
  {Jeanneret}, \citenamefont {Pushkin},\ and\ \citenamefont
  {Polin}}]{jeanneret2019confinement}%
  \BibitemOpen
  \bibfield  {author} {\bibinfo {author} {\bibfnamefont {R.}~\bibnamefont
  {Jeanneret}}, \bibinfo {author} {\bibfnamefont {D.~O.}\ \bibnamefont
  {Pushkin}},\ and\ \bibinfo {author} {\bibfnamefont {M.}~\bibnamefont
  {Polin}},\ }\bibfield  {title} {\bibinfo {title} {Confinement enhances the
  diversity of microbial flow fields},\ }\href
  {https://doi.org/10.1103/PhysRevLett.123.248102} {\bibfield  {journal}
  {\bibinfo  {journal} {Physical Review Letters}\ }\textbf {\bibinfo {volume}
  {123}},\ \bibinfo {pages} {248102} (\bibinfo {year} {2019})}\BibitemShut
  {NoStop}%
\bibitem [{\citenamefont {Liron}\ and\ \citenamefont
  {Mochon}(1976)}]{liron1976stokes}%
  \BibitemOpen
  \bibfield  {author} {\bibinfo {author} {\bibfnamefont {N.}~\bibnamefont
  {Liron}}\ and\ \bibinfo {author} {\bibfnamefont {S.}~\bibnamefont {Mochon}},\
  }\bibfield  {title} {\bibinfo {title} {Stokes flow for a stokeslet between
  two parallel flat plates},\ }\href {https://doi.org/10.1007/BF01535565}
  {\bibfield  {journal} {\bibinfo  {journal} {Journal of Engineering
  Mathematics}\ }\textbf {\bibinfo {volume} {10}},\ \bibinfo {pages} {287}
  (\bibinfo {year} {1976})}\BibitemShut {NoStop}%
\bibitem [{\citenamefont {Brotto}\ \emph {et~al.}(2013)\citenamefont {Brotto},
  \citenamefont {Caussin}, \citenamefont {Lauga},\ and\ \citenamefont
  {Bartolo}}]{brotto2013hydrodynamics}%
  \BibitemOpen
  \bibfield  {author} {\bibinfo {author} {\bibfnamefont {T.}~\bibnamefont
  {Brotto}}, \bibinfo {author} {\bibfnamefont {J.-B.}\ \bibnamefont {Caussin}},
  \bibinfo {author} {\bibfnamefont {E.}~\bibnamefont {Lauga}},\ and\ \bibinfo
  {author} {\bibfnamefont {D.}~\bibnamefont {Bartolo}},\ }\bibfield  {title}
  {\bibinfo {title} {Hydrodynamics of confined active fluids},\ }\href
  {https://doi.org/10.1103/PhysRevLett.110.038101} {\bibfield  {journal}
  {\bibinfo  {journal} {Physical Review Letters}\ }\textbf {\bibinfo {volume}
  {110}},\ \bibinfo {pages} {038101} (\bibinfo {year} {2013})}\BibitemShut
  {NoStop}%
\bibitem [{\citenamefont {Cui}\ \emph {et~al.}(2004)\citenamefont {Cui},
  \citenamefont {Diamant}, \citenamefont {Lin},\ and\ \citenamefont
  {Rice}}]{cui2004anomalous}%
  \BibitemOpen
  \bibfield  {author} {\bibinfo {author} {\bibfnamefont {B.}~\bibnamefont
  {Cui}}, \bibinfo {author} {\bibfnamefont {H.}~\bibnamefont {Diamant}},
  \bibinfo {author} {\bibfnamefont {B.}~\bibnamefont {Lin}},\ and\ \bibinfo
  {author} {\bibfnamefont {S.~A.}\ \bibnamefont {Rice}},\ }\bibfield  {title}
  {\bibinfo {title} {Anomalous hydrodynamic interaction in a
  quasi-two-dimensional suspension},\ }\href
  {https://doi.org/10.1103/PhysRevLett.92.258301} {\bibfield  {journal}
  {\bibinfo  {journal} {Physical Review Letters}\ }\textbf {\bibinfo {volume}
  {92}},\ \bibinfo {pages} {258301} (\bibinfo {year} {2004})}\BibitemShut
  {NoStop}%
\bibitem [{\citenamefont {Diamant}(2009)}]{diamant2009hydrodynamic}%
  \BibitemOpen
  \bibfield  {author} {\bibinfo {author} {\bibfnamefont {H.}~\bibnamefont
  {Diamant}},\ }\bibfield  {title} {\bibinfo {title} {Hydrodynamic interaction
  in confined geometries},\ }\href {https://doi.org/10.1143/JPSJ.78.041002}
  {\bibfield  {journal} {\bibinfo  {journal} {Journal of the Physical Society
  of Japan}\ }\textbf {\bibinfo {volume} {78}},\ \bibinfo {pages} {041002}
  (\bibinfo {year} {2009})}\BibitemShut {NoStop}%
\bibitem [{\citenamefont {Ryan}(2020)}]{ryan2020role}%
  \BibitemOpen
  \bibfield  {author} {\bibinfo {author} {\bibfnamefont {S.~D.}\ \bibnamefont
  {Ryan}},\ }\bibfield  {title} {\bibinfo {title} {Role of hydrodynamic
  interactions in chemotaxis of bacterial populations},\ }\href
  {https://doi.org/10.1088/1478-3975/ab57af} {\bibfield  {journal} {\bibinfo
  {journal} {Physical Biology}\ }\textbf {\bibinfo {volume} {17}},\ \bibinfo
  {pages} {016003} (\bibinfo {year} {2020})}\BibitemShut {NoStop}%
\bibitem [{\citenamefont {Jin}\ \emph {et~al.}(2021)\citenamefont {Jin},
  \citenamefont {Chen}, \citenamefont {Maass},\ and\ \citenamefont
  {Mathijssen}}]{jin2021collective}%
  \BibitemOpen
  \bibfield  {author} {\bibinfo {author} {\bibfnamefont {C.}~\bibnamefont
  {Jin}}, \bibinfo {author} {\bibfnamefont {Y.}~\bibnamefont {Chen}}, \bibinfo
  {author} {\bibfnamefont {C.~C.}\ \bibnamefont {Maass}},\ and\ \bibinfo
  {author} {\bibfnamefont {A.~J. T.~M.}\ \bibnamefont {Mathijssen}},\
  }\bibfield  {title} {\bibinfo {title} {Collective entrainment and confinement
  amplify transport by schooling microswimmers},\ }\href
  {https://doi.org/10.1103/PhysRevLett.127.088006} {\bibfield  {journal}
  {\bibinfo  {journal} {Physical Review Letters}\ }\textbf {\bibinfo {volume}
  {127}},\ \bibinfo {pages} {088006} (\bibinfo {year} {2021})}\BibitemShut
  {NoStop}%
\bibitem [{\citenamefont {Cheng Hou~Tsang}\ and\ \citenamefont
  {Kanso}(2014)}]{houtsang2014flagella}%
  \BibitemOpen
  \bibfield  {author} {\bibinfo {author} {\bibfnamefont {A.}~\bibnamefont
  {Cheng Hou~Tsang}}\ and\ \bibinfo {author} {\bibfnamefont {E.}~\bibnamefont
  {Kanso}},\ }\bibfield  {title} {\bibinfo {title} {Flagella-induced
  transitions in the collective behavior of confined microswimmers},\ }\href
  {https://doi.org/10.1103/PhysRevE.90.021001} {\bibfield  {journal} {\bibinfo
  {journal} {Phys. Rev. E}\ }\textbf {\bibinfo {volume} {90}},\ \bibinfo
  {pages} {021001} (\bibinfo {year} {2014})}\BibitemShut {NoStop}%
\bibitem [{\citenamefont {Delfau}\ \emph {et~al.}(2016)\citenamefont {Delfau},
  \citenamefont {Molina},\ and\ \citenamefont {Sano}}]{delfau2016collective}%
  \BibitemOpen
  \bibfield  {author} {\bibinfo {author} {\bibfnamefont {J.-B.}\ \bibnamefont
  {Delfau}}, \bibinfo {author} {\bibfnamefont {J.}~\bibnamefont {Molina}},\
  and\ \bibinfo {author} {\bibfnamefont {M.}~\bibnamefont {Sano}},\ }\bibfield
  {title} {\bibinfo {title} {Collective behavior of strongly confined
  suspensions of squirmers},\ }\href
  {https://doi.org/10.1209/0295-5075/114/24001} {\bibfield  {journal} {\bibinfo
   {journal} {EPL (Europhysics Letters)}\ }\textbf {\bibinfo {volume} {114}},\
  \bibinfo {pages} {24001} (\bibinfo {year} {2016})}\BibitemShut {NoStop}%
\bibitem [{\citenamefont {L{\'o}pez}\ \emph {et~al.}(2015)\citenamefont
  {L{\'o}pez}, \citenamefont {Gachelin}, \citenamefont {Douarche},
  \citenamefont {Auradou},\ and\ \citenamefont
  {Cl{\'e}ment}}]{lopez2015turning}%
  \BibitemOpen
  \bibfield  {author} {\bibinfo {author} {\bibfnamefont {H.~M.}\ \bibnamefont
  {L{\'o}pez}}, \bibinfo {author} {\bibfnamefont {J.}~\bibnamefont {Gachelin}},
  \bibinfo {author} {\bibfnamefont {C.}~\bibnamefont {Douarche}}, \bibinfo
  {author} {\bibfnamefont {H.}~\bibnamefont {Auradou}},\ and\ \bibinfo {author}
  {\bibfnamefont {E.}~\bibnamefont {Cl{\'e}ment}},\ }\bibfield  {title}
  {\bibinfo {title} {Turning bacteria suspensions into superfluids},\ }\href
  {https://doi.org/10.1103/PhysRevLett.115.028301} {\bibfield  {journal}
  {\bibinfo  {journal} {Physical Review Letters}\ }\textbf {\bibinfo {volume}
  {115}},\ \bibinfo {pages} {028301} (\bibinfo {year} {2015})}\BibitemShut
  {NoStop}%
\bibitem [{\citenamefont {Douarche}\ \emph {et~al.}(2009)\citenamefont
  {Douarche}, \citenamefont {Buguin}, \citenamefont {Salman},\ and\
  \citenamefont {Libchaber}}]{douarche2009coli}%
  \BibitemOpen
  \bibfield  {author} {\bibinfo {author} {\bibfnamefont {C.}~\bibnamefont
  {Douarche}}, \bibinfo {author} {\bibfnamefont {A.}~\bibnamefont {Buguin}},
  \bibinfo {author} {\bibfnamefont {H.}~\bibnamefont {Salman}},\ and\ \bibinfo
  {author} {\bibfnamefont {A.}~\bibnamefont {Libchaber}},\ }\bibfield  {title}
  {\bibinfo {title} {E. coli and oxygen: a motility transition},\ }\href
  {https://doi.org/10.1103/PhysRevLett.102.198101} {\bibfield  {journal}
  {\bibinfo  {journal} {Physical review letters}\ }\textbf {\bibinfo {volume}
  {102}},\ \bibinfo {pages} {198101} (\bibinfo {year} {2009})}\BibitemShut
  {NoStop}%
\bibitem [{\citenamefont {Sipos}\ \emph {et~al.}(2015)\citenamefont {Sipos},
  \citenamefont {Nagy}, \citenamefont {Di~Leonardo},\ and\ \citenamefont
  {Galajda}}]{sipos2015hydrodynamic}%
  \BibitemOpen
  \bibfield  {author} {\bibinfo {author} {\bibfnamefont {O.}~\bibnamefont
  {Sipos}}, \bibinfo {author} {\bibfnamefont {K.}~\bibnamefont {Nagy}},
  \bibinfo {author} {\bibfnamefont {R.}~\bibnamefont {Di~Leonardo}},\ and\
  \bibinfo {author} {\bibfnamefont {P.}~\bibnamefont {Galajda}},\ }\bibfield
  {title} {\bibinfo {title} {Hydrodynamic trapping of swimming bacteria by
  convex walls},\ }\href {https://doi.org/10.1103/PhysRevLett.114.258104}
  {\bibfield  {journal} {\bibinfo  {journal} {Phys. Rev. Lett.}\ }\textbf
  {\bibinfo {volume} {114}},\ \bibinfo {pages} {258104} (\bibinfo {year}
  {2015})}\BibitemShut {NoStop}%
\bibitem [{\citenamefont {Tinevez}\ \emph {et~al.}(2017)\citenamefont
  {Tinevez}, \citenamefont {Perry}, \citenamefont {Schindelin}, \citenamefont
  {Hoopes}, \citenamefont {Reynolds}, \citenamefont {Laplantine}, \citenamefont
  {Bednarek}, \citenamefont {Shorte},\ and\ \citenamefont
  {Eliceiri}}]{tinevez2017trackmate}%
  \BibitemOpen
  \bibfield  {author} {\bibinfo {author} {\bibfnamefont {J.-Y.}\ \bibnamefont
  {Tinevez}}, \bibinfo {author} {\bibfnamefont {N.}~\bibnamefont {Perry}},
  \bibinfo {author} {\bibfnamefont {J.}~\bibnamefont {Schindelin}}, \bibinfo
  {author} {\bibfnamefont {G.~M.}\ \bibnamefont {Hoopes}}, \bibinfo {author}
  {\bibfnamefont {G.~D.}\ \bibnamefont {Reynolds}}, \bibinfo {author}
  {\bibfnamefont {E.}~\bibnamefont {Laplantine}}, \bibinfo {author}
  {\bibfnamefont {S.~Y.}\ \bibnamefont {Bednarek}}, \bibinfo {author}
  {\bibfnamefont {S.~L.}\ \bibnamefont {Shorte}},\ and\ \bibinfo {author}
  {\bibfnamefont {K.~W.}\ \bibnamefont {Eliceiri}},\ }\bibfield  {title}
  {\bibinfo {title} {{TrackMate}: An open and extensible platform for
  single-particle tracking},\ }\href
  {https://doi.org/10.1016/j.ymeth.2016.09.016} {\bibfield  {journal} {\bibinfo
   {journal} {Methods}\ }\textbf {\bibinfo {volume} {115}},\ \bibinfo {pages}
  {80} (\bibinfo {year} {2017})}\BibitemShut {NoStop}%
\bibitem [{\citenamefont {Chopra}\ \emph {et~al.}(2022)\citenamefont {Chopra},
  \citenamefont {Quint}, \citenamefont {Gopinathan},\ and\ \citenamefont
  {Liu}}]{chopra2022geometric}%
  \BibitemOpen
  \bibfield  {author} {\bibinfo {author} {\bibfnamefont {P.}~\bibnamefont
  {Chopra}}, \bibinfo {author} {\bibfnamefont {D.}~\bibnamefont {Quint}},
  \bibinfo {author} {\bibfnamefont {A.}~\bibnamefont {Gopinathan}},\ and\
  \bibinfo {author} {\bibfnamefont {B.}~\bibnamefont {Liu}},\ }\bibfield
  {title} {\bibinfo {title} {Geometric effects induce anomalous size-dependent
  active transport in structured environments},\ }\href
  {https://doi.org/10.1103/PhysRevFluids.7.L071101} {\bibfield  {journal}
  {\bibinfo  {journal} {Phys. Rev. Fluids}\ }\textbf {\bibinfo {volume} {7}},\
  \bibinfo {pages} {L071101} (\bibinfo {year} {2022})}\BibitemShut {NoStop}%
\bibitem [{\citenamefont {Spagnolie}\ \emph {et~al.}(2015)\citenamefont
  {Spagnolie}, \citenamefont {Moreno-Flores}, \citenamefont {Bartolo},\ and\
  \citenamefont {Lauga}}]{spagnolie2015geometric}%
  \BibitemOpen
  \bibfield  {author} {\bibinfo {author} {\bibfnamefont {S.~E.}\ \bibnamefont
  {Spagnolie}}, \bibinfo {author} {\bibfnamefont {G.~R.}\ \bibnamefont
  {Moreno-Flores}}, \bibinfo {author} {\bibfnamefont {D.}~\bibnamefont
  {Bartolo}},\ and\ \bibinfo {author} {\bibfnamefont {E.}~\bibnamefont
  {Lauga}},\ }\bibfield  {title} {\bibinfo {title} {Geometric capture and
  escape of a microswimmer colliding with an obstacle},\ }\href
  {https://doi.org/10.1039/C4SM02785J} {\bibfield  {journal} {\bibinfo
  {journal} {Soft Matter}\ }\textbf {\bibinfo {volume} {11}},\ \bibinfo {pages}
  {3396} (\bibinfo {year} {2015})}\BibitemShut {NoStop}%
\bibitem [{\citenamefont {Takagi}\ \emph {et~al.}(2014)\citenamefont {Takagi},
  \citenamefont {Palacci}, \citenamefont {Braunschweig}, \citenamefont
  {Shelley},\ and\ \citenamefont {Zhang}}]{takagi2014hydrodynamic}%
  \BibitemOpen
  \bibfield  {author} {\bibinfo {author} {\bibfnamefont {D.}~\bibnamefont
  {Takagi}}, \bibinfo {author} {\bibfnamefont {J.}~\bibnamefont {Palacci}},
  \bibinfo {author} {\bibfnamefont {A.~B.}\ \bibnamefont {Braunschweig}},
  \bibinfo {author} {\bibfnamefont {M.~J.}\ \bibnamefont {Shelley}},\ and\
  \bibinfo {author} {\bibfnamefont {J.}~\bibnamefont {Zhang}},\ }\bibfield
  {title} {\bibinfo {title} {Hydrodynamic capture of microswimmers into
  sphere-bound orbits},\ }\href {https://doi.org/10.1039/C3SM52815D} {\bibfield
   {journal} {\bibinfo  {journal} {Soft Matter}\ }\textbf {\bibinfo {volume}
  {10}},\ \bibinfo {pages} {1784} (\bibinfo {year} {2014})}\BibitemShut
  {NoStop}%
\bibitem [{\citenamefont {Lauga}\ and\ \citenamefont
  {Powers}(2009)}]{lauga2009}%
  \BibitemOpen
  \bibfield  {author} {\bibinfo {author} {\bibfnamefont {E.}~\bibnamefont
  {Lauga}}\ and\ \bibinfo {author} {\bibfnamefont {T.~R.}\ \bibnamefont
  {Powers}},\ }\bibfield  {title} {\bibinfo {title} {The hydrodynamics of
  swimming microorganisms},\ }\href
  {https://doi.org/10.1088/0034-4885/72/9/096601} {\bibfield  {journal}
  {\bibinfo  {journal} {Reports on Progress in Physics}\ }\textbf {\bibinfo
  {volume} {72}},\ \bibinfo {pages} {096601} (\bibinfo {year}
  {2009})}\BibitemShut {NoStop}%
\bibitem [{\citenamefont {Swiecicki}\ \emph {et~al.}(2013)\citenamefont
  {Swiecicki}, \citenamefont {Sliusarenko},\ and\ \citenamefont
  {Weibel}}]{swiecicki2013twodimension}%
  \BibitemOpen
  \bibfield  {author} {\bibinfo {author} {\bibfnamefont {J.-M.}\ \bibnamefont
  {Swiecicki}}, \bibinfo {author} {\bibfnamefont {O.}~\bibnamefont
  {Sliusarenko}},\ and\ \bibinfo {author} {\bibfnamefont {D.~B.}\ \bibnamefont
  {Weibel}},\ }\bibfield  {title} {\bibinfo {title} {{From swimming to
  swarming: Escherichia coli cell motility in two-dimensions}},\ }\href
  {https://doi.org/10.1039/c3ib40130h} {\bibfield  {journal} {\bibinfo
  {journal} {Integrative Biology}\ }\textbf {\bibinfo {volume} {5}},\ \bibinfo
  {pages} {1490} (\bibinfo {year} {2013})}\BibitemShut {NoStop}%
\bibitem [{\citenamefont {Hecht}(2012)}]{FreeFEM}%
  \BibitemOpen
  \bibfield  {author} {\bibinfo {author} {\bibfnamefont {F.}~\bibnamefont
  {Hecht}},\ }\bibfield  {title} {\bibinfo {title} {New development in
  {FreeFem}++},\ }\href {https://freefem.org/} {\bibfield  {journal} {\bibinfo
  {journal} {Journal of Numerical Mathematics}\ }\textbf {\bibinfo {volume}
  {20}},\ \bibinfo {pages} {251} (\bibinfo {year} {2012})}\BibitemShut
  {NoStop}%
\bibitem [{\citenamefont {Drescher}\ \emph {et~al.}(2011)\citenamefont
  {Drescher}, \citenamefont {Dunkel}, \citenamefont {Cisneros}, \citenamefont
  {Ganguly},\ and\ \citenamefont {Goldstein}}]{drescher2011fluid}%
  \BibitemOpen
  \bibfield  {author} {\bibinfo {author} {\bibfnamefont {K.}~\bibnamefont
  {Drescher}}, \bibinfo {author} {\bibfnamefont {J.}~\bibnamefont {Dunkel}},
  \bibinfo {author} {\bibfnamefont {L.~H.}\ \bibnamefont {Cisneros}}, \bibinfo
  {author} {\bibfnamefont {S.}~\bibnamefont {Ganguly}},\ and\ \bibinfo {author}
  {\bibfnamefont {R.~E.}\ \bibnamefont {Goldstein}},\ }\bibfield  {title}
  {\bibinfo {title} {Fluid dynamics and noise in bacterial cell \textendash{}
  cell and cell \textendash{} surface scattering},\ }\href
  {https://doi.org/10.1073/pnas.1019079108} {\bibfield  {journal} {\bibinfo
  {journal} {Proceedings of the National Academy of Sciences}\ }\textbf
  {\bibinfo {volume} {108}},\ \bibinfo {pages} {10940} (\bibinfo {year}
  {2011})}\BibitemShut {NoStop}%
\bibitem [{\citenamefont {Denissenko}\ \emph {et~al.}(2012)\citenamefont
  {Denissenko}, \citenamefont {Kantsler}, \citenamefont {Smith},\ and\
  \citenamefont {Kirkman-Brown}}]{denissenko2012human}%
  \BibitemOpen
  \bibfield  {author} {\bibinfo {author} {\bibfnamefont {P.}~\bibnamefont
  {Denissenko}}, \bibinfo {author} {\bibfnamefont {V.}~\bibnamefont
  {Kantsler}}, \bibinfo {author} {\bibfnamefont {D.~J.}\ \bibnamefont
  {Smith}},\ and\ \bibinfo {author} {\bibfnamefont {J.}~\bibnamefont
  {Kirkman-Brown}},\ }\bibfield  {title} {\bibinfo {title} {Human spermatozoa
  migration in microchannels reveals boundary-following navigation},\ }\href
  {https://doi.org/10.1073/pnas.1202934109} {\bibfield  {journal} {\bibinfo
  {journal} {Proceedings of the National Academy of Sciences}\ }\textbf
  {\bibinfo {volume} {109}},\ \bibinfo {pages} {8007} (\bibinfo {year}
  {2012})}\BibitemShut {NoStop}%
\bibitem [{\citenamefont {Ishimoto}\ \emph {et~al.}(2016)\citenamefont
  {Ishimoto}, \citenamefont {Cosson},\ and\ \citenamefont
  {Gaffney}}]{ishimoto2016simulation}%
  \BibitemOpen
  \bibfield  {author} {\bibinfo {author} {\bibfnamefont {K.}~\bibnamefont
  {Ishimoto}}, \bibinfo {author} {\bibfnamefont {J.}~\bibnamefont {Cosson}},\
  and\ \bibinfo {author} {\bibfnamefont {E.~A.}\ \bibnamefont {Gaffney}},\
  }\bibfield  {title} {\bibinfo {title} {A simulation study of sperm motility
  hydrodynamics near fish eggs and spheres},\ }\href
  {https://doi.org/10.1016/j.jtbi.2015.10.013} {\bibfield  {journal} {\bibinfo
  {journal} {Journal of Theoretical Biology}\ }\textbf {\bibinfo {volume}
  {389}},\ \bibinfo {pages} {187} (\bibinfo {year} {2016})}\BibitemShut
  {NoStop}%
\bibitem [{\citenamefont {Elgeti}\ and\ \citenamefont
  {Gompper}(2013)}]{elgeti2013wall}%
  \BibitemOpen
  \bibfield  {author} {\bibinfo {author} {\bibfnamefont {J.}~\bibnamefont
  {Elgeti}}\ and\ \bibinfo {author} {\bibfnamefont {G.}~\bibnamefont
  {Gompper}},\ }\bibfield  {title} {\bibinfo {title} {Wall accumulation of
  self-propelled spheres},\ }\href
  {https://doi.org/10.1209/0295-5075/101/48003} {\bibfield  {journal} {\bibinfo
   {journal} {EPL (Europhysics Letters)}\ }\textbf {\bibinfo {volume} {101}},\
  \bibinfo {pages} {48003} (\bibinfo {year} {2013})}\BibitemShut {NoStop}%
\bibitem [{\citenamefont {Ishikawa}\ \emph {et~al.}(2006)\citenamefont
  {Ishikawa}, \citenamefont {Simmonds},\ and\ \citenamefont
  {Pedley}}]{ishikawa2006hydrodynamic}%
  \BibitemOpen
  \bibfield  {author} {\bibinfo {author} {\bibfnamefont {T.}~\bibnamefont
  {Ishikawa}}, \bibinfo {author} {\bibfnamefont {M.}~\bibnamefont {Simmonds}},\
  and\ \bibinfo {author} {\bibfnamefont {T.~J.}\ \bibnamefont {Pedley}},\
  }\bibfield  {title} {\bibinfo {title} {Hydrodynamic interaction of two
  swimming model micro-organisms},\ }\href
  {https://doi.org/10.1017/S0022112006002631} {\bibfield  {journal} {\bibinfo
  {journal} {Journal of Fluid Mechanics}\ }\textbf {\bibinfo {volume} {568}},\
  \bibinfo {pages} {119} (\bibinfo {year} {2006})}\BibitemShut {NoStop}%
\bibitem [{\citenamefont {Zhu}\ \emph {et~al.}(2013)\citenamefont {Zhu},
  \citenamefont {Lauga},\ and\ \citenamefont {Brandt}}]{zhu2013low}%
  \BibitemOpen
  \bibfield  {author} {\bibinfo {author} {\bibfnamefont {L.}~\bibnamefont
  {Zhu}}, \bibinfo {author} {\bibfnamefont {E.}~\bibnamefont {Lauga}},\ and\
  \bibinfo {author} {\bibfnamefont {L.}~\bibnamefont {Brandt}},\ }\bibfield
  {title} {\bibinfo {title} {Low-{R}eynolds-number swimming in a capillary
  tube},\ }\href {https://doi.org/10.1017/jfm.2013.225} {\bibfield  {journal}
  {\bibinfo  {journal} {Journal of Fluid Mechanics}\ }\textbf {\bibinfo
  {volume} {726}},\ \bibinfo {pages} {285} (\bibinfo {year}
  {2013})}\BibitemShut {NoStop}%
\bibitem [{\citenamefont {Shum}\ and\ \citenamefont
  {Gaffney}(2015{\natexlab{a}})}]{shum2015hydrodynamicA}%
  \BibitemOpen
  \bibfield  {author} {\bibinfo {author} {\bibfnamefont {H.}~\bibnamefont
  {Shum}}\ and\ \bibinfo {author} {\bibfnamefont {E.~A.}\ \bibnamefont
  {Gaffney}},\ }\bibfield  {title} {\bibinfo {title} {Hydrodynamic analysis of
  flagellated bacteria swimming in corners of rectangular channels},\ }\href
  {https://doi.org/10.1103/PhysRevE.92.063016} {\bibfield  {journal} {\bibinfo
  {journal} {Physical review E}\ }\textbf {\bibinfo {volume} {92}},\ \bibinfo
  {pages} {063016} (\bibinfo {year} {2015}{\natexlab{a}})}\BibitemShut
  {NoStop}%
\bibitem [{\citenamefont {Shum}\ and\ \citenamefont
  {Gaffney}(2015{\natexlab{b}})}]{shum2015hydrodynamicB}%
  \BibitemOpen
  \bibfield  {author} {\bibinfo {author} {\bibfnamefont {H.}~\bibnamefont
  {Shum}}\ and\ \bibinfo {author} {\bibfnamefont {E.~A.}\ \bibnamefont
  {Gaffney}},\ }\bibfield  {title} {\bibinfo {title} {Hydrodynamic analysis of
  flagellated bacteria swimming near one and between two no-slip plane
  boundaries},\ }\href {https://doi.org/10.1103/PhysRevE.91.033012} {\bibfield
  {journal} {\bibinfo  {journal} {Physical Review E}\ }\textbf {\bibinfo
  {volume} {91}},\ \bibinfo {pages} {033012} (\bibinfo {year}
  {2015}{\natexlab{b}})}\BibitemShut {NoStop}%
\bibitem [{\citenamefont {Molina}\ \emph {et~al.}(2013)\citenamefont {Molina},
  \citenamefont {Nakayama},\ and\ \citenamefont
  {Yamamoto}}]{molina2013hydrodynamic}%
  \BibitemOpen
  \bibfield  {author} {\bibinfo {author} {\bibfnamefont {J.~J.}\ \bibnamefont
  {Molina}}, \bibinfo {author} {\bibfnamefont {Y.}~\bibnamefont {Nakayama}},\
  and\ \bibinfo {author} {\bibfnamefont {R.}~\bibnamefont {Yamamoto}},\
  }\bibfield  {title} {\bibinfo {title} {Hydrodynamic interactions of
  self-propelled swimmers},\ }\href {https://doi.org/10.1039/C3SM00140G}
  {\bibfield  {journal} {\bibinfo  {journal} {Soft Matter}\ }\textbf {\bibinfo
  {volume} {9}},\ \bibinfo {pages} {4923} (\bibinfo {year} {2013})}\BibitemShut
  {NoStop}%
\bibitem [{\citenamefont {Oyama}\ \emph {et~al.}(2017)\citenamefont {Oyama},
  \citenamefont {Molina},\ and\ \citenamefont
  {Yamamoto}}]{oyama2017simulations}%
  \BibitemOpen
  \bibfield  {author} {\bibinfo {author} {\bibfnamefont {N.}~\bibnamefont
  {Oyama}}, \bibinfo {author} {\bibfnamefont {J.~J.}\ \bibnamefont {Molina}},\
  and\ \bibinfo {author} {\bibfnamefont {R.}~\bibnamefont {Yamamoto}},\
  }\bibfield  {title} {\bibinfo {title} {Simulations of model microswimmers
  with fully resolved hydrodynamics},\ }\href
  {https://doi.org/10.7566/JPSJ.86.101008} {\bibfield  {journal} {\bibinfo
  {journal} {Journal of the Physical Society of Japan}\ }\textbf {\bibinfo
  {volume} {86}},\ \bibinfo {pages} {101008} (\bibinfo {year}
  {2017})}\BibitemShut {NoStop}%
\bibitem [{\citenamefont {Davies~Wykes}\ \emph {et~al.}(2017)\citenamefont
  {Davies~Wykes}, \citenamefont {Zhong}, \citenamefont {Tong}, \citenamefont
  {Adachi}, \citenamefont {Liu}, \citenamefont {Ristroph}, \citenamefont
  {Ward}, \citenamefont {Shelley},\ and\ \citenamefont
  {Zhang}}]{davieswykes2017guiding}%
  \BibitemOpen
  \bibfield  {author} {\bibinfo {author} {\bibfnamefont {M.~S.}\ \bibnamefont
  {Davies~Wykes}}, \bibinfo {author} {\bibfnamefont {X.}~\bibnamefont {Zhong}},
  \bibinfo {author} {\bibfnamefont {J.}~\bibnamefont {Tong}}, \bibinfo {author}
  {\bibfnamefont {T.}~\bibnamefont {Adachi}}, \bibinfo {author} {\bibfnamefont
  {Y.}~\bibnamefont {Liu}}, \bibinfo {author} {\bibfnamefont {L.}~\bibnamefont
  {Ristroph}}, \bibinfo {author} {\bibfnamefont {M.~D.}\ \bibnamefont {Ward}},
  \bibinfo {author} {\bibfnamefont {M.~J.}\ \bibnamefont {Shelley}},\ and\
  \bibinfo {author} {\bibfnamefont {J.}~\bibnamefont {Zhang}},\ }\bibfield
  {title} {\bibinfo {title} {Guiding microscale swimmers using teardrop-shaped
  posts},\ }\href {https://doi.org/10.1039/C7SM00203C} {\bibfield  {journal}
  {\bibinfo  {journal} {Soft Matter}\ }\textbf {\bibinfo {volume} {13}},\
  \bibinfo {pages} {4681} (\bibinfo {year} {2017})}\BibitemShut {NoStop}%
\bibitem [{\citenamefont {Li}\ \emph {et~al.}(2017)\citenamefont {Li},
  \citenamefont {Zhai}, \citenamefont {Sanchez}, \citenamefont {Kearns},\ and\
  \citenamefont {Wu}}]{li2017noncontact}%
  \BibitemOpen
  \bibfield  {author} {\bibinfo {author} {\bibfnamefont {Y.}~\bibnamefont
  {Li}}, \bibinfo {author} {\bibfnamefont {H.}~\bibnamefont {Zhai}}, \bibinfo
  {author} {\bibfnamefont {S.}~\bibnamefont {Sanchez}}, \bibinfo {author}
  {\bibfnamefont {D.~B.}\ \bibnamefont {Kearns}},\ and\ \bibinfo {author}
  {\bibfnamefont {Y.}~\bibnamefont {Wu}},\ }\bibfield  {title} {\bibinfo
  {title} {Noncontact cohesive swimming of bacteria in two-dimensional liquid
  films},\ }\href {https://doi.org/10.1103/PhysRevLett.119.018101} {\bibfield
  {journal} {\bibinfo  {journal} {Physical Review Letters}\ }\textbf {\bibinfo
  {volume} {119}},\ \bibinfo {pages} {018101} (\bibinfo {year}
  {2017})}\BibitemShut {NoStop}%
\bibitem [{\citenamefont {Nishiguchi}\ and\ \citenamefont
  {Sano}(2015)}]{nishiguchi2015mesoscopic}%
  \BibitemOpen
  \bibfield  {author} {\bibinfo {author} {\bibfnamefont {D.}~\bibnamefont
  {Nishiguchi}}\ and\ \bibinfo {author} {\bibfnamefont {M.}~\bibnamefont
  {Sano}},\ }\bibfield  {title} {\bibinfo {title} {Mesoscopic turbulence and
  local order in janus particles self-propelling under an ac electric field},\
  }\href {https://doi.org/10.1103/PhysRevE.92.052309} {\bibfield  {journal}
  {\bibinfo  {journal} {Physical Review E}\ }\textbf {\bibinfo {volume} {92}},\
  \bibinfo {pages} {052309} (\bibinfo {year} {2015})}\BibitemShut {NoStop}%
\bibitem [{\citenamefont {{Nishiguchi}}\ \emph {et~al.}(2018)\citenamefont
  {{Nishiguchi}}, \citenamefont {{Iwasawa}}, \citenamefont {{Jiang}},\ and\
  \citenamefont {{Sano}}}]{nishiguchi2018flagellar}%
  \BibitemOpen
  \bibfield  {author} {\bibinfo {author} {\bibfnamefont {D.}~\bibnamefont
  {{Nishiguchi}}}, \bibinfo {author} {\bibfnamefont {J.}~\bibnamefont
  {{Iwasawa}}}, \bibinfo {author} {\bibfnamefont {H.-R.}\ \bibnamefont
  {{Jiang}}},\ and\ \bibinfo {author} {\bibfnamefont {M.}~\bibnamefont
  {{Sano}}},\ }\bibfield  {title} {\bibinfo {title} {{Flagellar dynamics of
  chains of active Janus particles fueled by an AC electric field}},\ }\href
  {https://doi.org/10.1088/1367-2630/aa9b48} {\bibfield  {journal} {\bibinfo
  {journal} {New Journal of Physics}\ }\textbf {\bibinfo {volume} {20}},\
  \bibinfo {eid} {015002} (\bibinfo {year} {2018})}\BibitemShut {NoStop}%
\bibitem [{\citenamefont {Poncet}\ \emph {et~al.}(2021)\citenamefont {Poncet},
  \citenamefont {B\'enichou}, \citenamefont {D\'emery},\ and\ \citenamefont
  {Nishiguchi}}]{poncet2020pair}%
  \BibitemOpen
  \bibfield  {author} {\bibinfo {author} {\bibfnamefont {A.}~\bibnamefont
  {Poncet}}, \bibinfo {author} {\bibfnamefont {O.}~\bibnamefont {B\'enichou}},
  \bibinfo {author} {\bibfnamefont {V.}~\bibnamefont {D\'emery}},\ and\
  \bibinfo {author} {\bibfnamefont {D.}~\bibnamefont {Nishiguchi}},\ }\bibfield
   {title} {\bibinfo {title} {Pair correlation of dilute active {B}rownian
  particles: From low-activity dipolar correction to high-activity algebraic
  depletion wings},\ }\href {https://doi.org/10.1103/PhysRevE.103.012605}
  {\bibfield  {journal} {\bibinfo  {journal} {Physical Review E}\ }\textbf
  {\bibinfo {volume} {103}},\ \bibinfo {pages} {012605} (\bibinfo {year}
  {2021})}\BibitemShut {NoStop}%
\bibitem [{\citenamefont {Iwasawa}\ \emph {et~al.}(2021)\citenamefont
  {Iwasawa}, \citenamefont {Nishiguchi},\ and\ \citenamefont
  {Sano}}]{iwasawa2021algebraic}%
  \BibitemOpen
  \bibfield  {author} {\bibinfo {author} {\bibfnamefont {J.}~\bibnamefont
  {Iwasawa}}, \bibinfo {author} {\bibfnamefont {D.}~\bibnamefont
  {Nishiguchi}},\ and\ \bibinfo {author} {\bibfnamefont {M.}~\bibnamefont
  {Sano}},\ }\bibfield  {title} {\bibinfo {title} {Algebraic correlations and
  anomalous fluctuations in ordered flocks of {J}anus particles fueled by an
  {AC} electric field},\ }\href
  {https://doi.org/10.1103/PhysRevResearch.3.043104} {\bibfield  {journal}
  {\bibinfo  {journal} {Physical Review Research}\ }\textbf {\bibinfo {volume}
  {3}},\ \bibinfo {pages} {043104} (\bibinfo {year} {2021})}\BibitemShut
  {NoStop}%
\end{thebibliography}%

\end{document}